# Toward Reliable Modeling of S-Nitrosothiol Chemistry: Structure and Properties of Methyl Thionitrite ($CH_3SNO$), an S-Nitrosocysteine Model

Dmitry G. Khomyakov and Qadir K. Timerghazin[a)]

*Department of Chemistry, Marquette University, Milwaukee, Wisconsin, 53201-1881, USA*

Methyl thionitrite $CH_3SNO$ is an important model of S-nitrosated cysteine aminoacid residue (CysNO), a ubiquitous biological S-nitrosothiol (RSNO) involved in numerous physiological processes. As such, $CH_3SNO$ can provide insights into the intrinsic properties of the –SNO group in CysNO, in particular, its weak and labile S–N bond. Here, we report an *ab initio* computational investigation of the structure and properties of $CH_3SNO$ using a composite Feller-Peterson-Dixon (FPD) scheme based on the explicitly-correlated coupled cluster with single, double, and perturbative excitations calculations extrapolated to the complete basis set limit, CCSD(T)-F12/CBS, with a number of additive corrections for the effects of quadruple excitations, core-valence correlation, scalar-relativistic and spin-orbit effects, as well as harmonic zero-point vibrational energy (ZPE) with an anharmonicity correction. These calculations suggest that the S–N bond in $CH_3SNO$ is significantly elongated (1.814 Å), has low stretching frequency and dissociation energy values, $\nu_{S–N} = 387\ cm^{-1}$ and $D_0 = 32.4$ kcal/mol. At the same time, the S–N bond has a sizable rotation barrier, $\Delta E_0^{\neq} = 12.7$ kcal/mol, so $CH_3SNO$ exists as cis- or trans-conformer, the latter slightly higher in energy, $\Delta E_0 = 1.2$ kcal/mol. The S–N bond properties are consistent with the antagonistic nature of $CH_3SNO$, whose resonance representation requires two chemically opposite (antagonistic) resonance structures, $CH_3$-$S^+$=N–$O^-$ and $CH_3$-$S^-$ / $NO^+$, which can be probed using external electric fields and quantified using the natural resonance theory (NRT) approach. The calculated S–N bond properties slowly converge with the level of correlation treatment, with the recently developed distinguished cluster with single and double excitations approximation (DCSD-F12) performing significantly better than the coupled cluster with single and double excitations method (CCSD-F12), although still inferior to the CCSD(T)-F12 method that includes perturbative triple excitations. Double-hybrid density functional theory (DFT) calculations with mPW2PLYPD/def2-TZVPPD reproduce well the geometry, vibrational frequencies, and the S–N bond rotational barrier in $CH_3SNO$, while hybrid DFT calculations with PBE0/def2-TZVPPD give a better S–N bond dissociation energy.

## I. INTRODUCTION

S-Nitrosothiols (RSNOs) are ubiquitous biological derivatives of nitric oxide, a major gasotransmitter.[1-4] Reversible S-nitrosation of the thiol functional group of cysteine (Cys) aminoacid residues in proteins leading to formation of S-nitrosated CysNO is an important post-translational modification involved in numerous biological processes in a wide variety of organisms.[5-13] Thousands of proteins have been reported to undergo S-nitrosation *in vivo*,[6,8,10] and while numerous factors point out to enzymatic control of biological RSNO reactions,[9,14,15] the underlying chemistry is still poorly understood.[4]

RSNOs in general are sensitive to light, rapidly decompose in the presence of metal ions, and have low thermal stability with respect to homolytic cleavage of the weak S–N bond with dissociation energy of only ~30 kcal/mol.[16-18] However, the molecular environment can modulate the stability of CysNO in an exceptionally wide range,[19] likely through the influence of proximal charges.[20,21] For instance, the room-temperature *in vitro* half-life ($t_{1/2}$) of CysNO as a free aminoacid is only 0.5 h;

however, it increases to 13.6 h for CysNO residue in a protein albumin, and to 40 h for CysNO within a tripepdide glutathione, while N-acetylated CysNO has $t_{1/2}$ = 500 h.[19] Understanding the molecular origins of this remarkable variation of properties as well as the enzymatic mechanisms underlying the biological reactions involving CysNO requires reliable information on the intrinsic properties of the –SNO group in CysNO, in particular the strength of its S–N bond. For instance, reliable data on the S–N bond dissociation energy in CysNO is essential for thermochemical analysis of the biological reactions of nitric oxide and its derivatives.[22] In this context, accurate electronic structure calculations of the relevant CysNO models are certainly of significant value.

Unfortunately, RSNOs do not lend themselves to reliable computational modeling of their properties—in particular, the S–N bond dissociation energy, early computational predictions of which ranged from 15 to 35 kcal/mol depending on the method.[23,24] Later systematic high-level *ab inito* investigations focused on thionitrous acid HSNO,[25-30] the smallest RSNO model molecule that, due to its small size, could be feasibly investigated with the state-of-the-art electronic structure methods. Although HSNO has been recently proposed to be an important biological RSNO in its own right,[31-33] accurate high-level computational—and also recently reported[30] experimental—data on HSNO may have only limited utility for understanding the chemistry of the most biologically abundant cysteine-based RSNOs, because the nature of substituent R may significantly affect the properties of the –SNO group.[19,34]

The smallest aliphatic RSNO, methyl thionitrite $CH_3SNO$, is a far better model of S-nitrosated cysteine sidechain, a primary aliphatic RSNO. As the prior computational investigations of HSNO have demonstrated,[25,26,28] the computed –SNO group properties show excruciatingly slow convergence with respect to the level of electron correlation treatment and the one-electron basis set size, necessitating using coupled cluster methods with electronic excitations up to quadruple level, and one-electron basis sets up to quintuple- (5Z) and sextuple-zeta (6Z) quality. This made accurate *ab initio* calculations of even marginally larger $CH_3SNO$ molecule extremely challenging computationally.

Fortunately, recently developed explicitly-correlated (F12) coupled cluster methods now allow to significantly alleviate the one-electron basis set convergence problems.[35,36] The F12 methods, which explicitly include inter-electron distance through an exponential correlation factor $F_{12} = \mathrm{e}^{-\gamma r_{12}}$, demonstrate much faster convergence to the complete basis set limit (CBS), so sextuple-zeta (6Z) quality results can be achieved with quadruple-zeta (QZ) basis set, and quintuple-zeta (5Z) quality—with triple-zeta (TZ) basis set, the "two-zeta gain rule."[37] Promising preliminary data on the explicitly-correlated coupled cluster calculations of HSNO[27,38] suggests that this rule also holds in the challenging case of the –SNO group.

Therefore, in this work we report accurate *ab initio* investigation of $CH_3SNO$ molecule with Feller-Peterson-Dixon (FPD) approach,[39-41] based on the CBS-extrapolated explicitly-correlated coupled cluster with single, double, and perturbative triple excitations, CCSD(T)-F12, calculations with a number of additive corrections for the effects of quadruple excitations, core-valence correlation, scalar-relativistic and spin-orbit effects, as well as harmonic zero-point vibrational energy (ZPE) with an anharmonicity correction. In particular, we focus on the S–N bond properties—its length, vibrational frequency, dissociation energy, and the rotational barrier; we also examine the convergence of these properties with the level of the coupled-cluster electron correlation treatment, including recently developed distinguished cluster approach, and discuss the unusual antagonistic nature of the S–N bond probed through external electric field effect calculations. We also report a limited assessment of the performance of several commonly used density functional theory (DFT) methods against the high-level FPD data.

## II. COMPUTATIONAL DETAILS

*Ab initio* electronic structure calculations were performed using Molpro 2015.1[42] program package and MRCC code interfaced with the CFOUR program[43,44] that was used for CCSDT(Q) calculations. Full geometry optimizations of cis- and trans-$CH_3SNO$, and $CH_3SNO$ cis-trans isomerization transition state ($TS_{c\text{-}t}$) were performed using frozen-core fixed amplitude explicitly-correlated (F12a) coupled-cluster with single, double, and perturbative connected triple excitations, CCSD(T)-F12,[35,36] with the F12-optimized cc-pV$n$Z-F12 ($n$ = D, T, Q; further referred to as VDZ-F12, VTZ-F12 and VQZ-F12, respectively) basis sets,[45] and the nature of all stationary points was confirmed by vibrational frequency calculations using Hessian matrix evaluated numerically. In addition, performance of coupled-cluster with single and double excitations, CCSD-F12,[35,36] as well as conventional and explicitly-correlated distinguishable cluster with single and double excitations. DCSD and DCSD-F12[46-48] has been investigated.

The complete basis set (CBS) extrapolations were performed based on the CCSD(T)-F12/VTZ-F12 ($n$ = 3) and CCSD(T)-F12/VQZ-F12 ($n$ = 4) results using a two-point formula[49]

$$E(n) = E_{CBS(T-Q)} + \frac{B}{n^3}, \tag{1}$$

which has been applied directly to estimate the geometric parameters at the CBS limit. The corrections for the coupled cluster quadruple excitations, Δ(Q), were evaluated by initial geometry optimization at the CCSD(T)/cc-pV(D+d)Z level followed by the S–N bond relaxation with CCSDT(Q)/cc-pV(D+d)Z level. Core-valence corrections (ΔCV) to the geometric parameters were estimated from all-electron (excluding the 1s-electrons of S atom) and frozen-core CCSD(T)-F12 geometry optimizations with weighted CV basis set cc-pCVTZ-F12.[50] Scalar-relativistic corrections ΔSR were evaluated in a similar manner at the CCSD(T)/cc-pVQZ-DK level using Douglas-Kroll-Hess method,[51,52] as implemented in Molpro.

Calculations of the energetic parameters, *D*(S–N), Δ*E*(cis-trans), and $\Delta E^{\neq}$(cis-trans), were based on the CCSD(T)-F12/CBS electronic energies and harmonic zero-point vibrational energies ($ZPE_{harm}$) calculated with two-point $CBS_{(T\text{-}Q)}$ extrapolation formula (1), using the CCSD(T)-F12a/VTZ-F12 and CCSD(T)-F12a/VQZ-F12 data. We compared the $CBS_{(T\text{-}Q)}$ values of $D_e$(S–N), obtained with the two-point formula (1) and a Schwenke-type extrapolation scheme (2):[53]

$$E_{CBS} = (E_{large} - E_{small})F + E_{small} \tag{2}$$

where $E_{large}$ and $E_{small}$ correspond to the electronic energies, obtained with VTZ-F12 and VQZ-F12 basis sets, correspondingly. Hill et al.[54] proposed the *F* value in (2) as 1.363388 for the CCSD-F12 with the VTZ-F12/VQZ-F12 basis sets, and 1.769474 for the perturbative (T) contribution, intending to alleviate slower convergence of the (T) component on the one-electron basis set size. In this work, both schemes produced nearly identical results (Table S1 in Supporting Information).

The Δ(Q) corrections for $NO^{\bullet}$ and $CH_3S^{\bullet}$ radicals were obtained from single-point calculations with cc-pV(D+d) basis set, whereas the S–N bond length in $CH_3SNO$ was optimized at the CCSDT(Q) level. ΔCV and ΔSR corrections were calculated with full geometry optimization of cis-$CH_3SNO$ and the radical fragments. Spin-orbit coupling correction for $CH_3S^{\bullet}$ radical (ΔSO) was calculated with the Breit-Pauli operator,[55] with multireference configuration interaction (MRCI) method[56,57] and aug-cc-pV(Q+d)Z basis set,[58,59] as implemented in Molpro.

All density functional theory (DFT) calculations were performed with Gaussian 09,[60] with 'UltraFine' settings for the integration grid (99 radial shells, 590 angular points per shell), with a number of hybrid functionals including B3LYP,[61-63] PBE0,[64-66] PBE0 with empirical dispersion correction,[67] PBE0-GD3, PBE0 with increased exact exchange contribution, PBE0-1/3,[68] ωB97XD,[69] as well double-hybrid B2PLYP[70] and mPW2PLYP[71] functionals and their dispersion-corrected versions B2PLYPD and mPW2PLYPD,[71]. DFT calculations used def2-SV(P)+d (with a tight d-type basis function for sulfur from aug-cc-pV(D+d)Z basis set[59]) and def2-TZVPPD basis sets by Weigend and Ahlrichs.[72,73] Natural Resonance Theory (NRT) [74,75] calculations were performed with the NBO 5.9 code[76], using PBE0/def2-TZVPPD density matrices. The contributions of the three resonance structures (***S***, ***D*** and ***I*** , Scheme 1) were determined from optimized multi-reference weights, as implemented in the multi-reference NRT[74] procedure.

The anharmonic contributions to vibrational frequencies and ZPE used in the FPD scheme were calculated with second-order perturbation theory (PT2) approach,[77] at the mPW2PLYPD/def2-TZVPPD and PBE0-GD3/def2-TZVPPD levels. The solvent effects on the FPD energetic parameters were evaluated from DFT calculations with def2-TZVPPD basis set and the integral equation formalism polarizable continuum model (IEFPCM),[78] with parameters for water (ε=78.36) and diethylether (ε=4.24).

## III. STRUCTURE OF $CH_3SNO$

Accurate *ab initio* calculations of the RSNO properties—in particular, those of the S–N bond—are challenging due to slow convergence with respect to both one-electron basis set size and the level of electron correlation treatment.[25,28] For instance, in our earlier study of HSNO, improving the basis set from double- to quintuple-zeta lead to >0.05 Å shortening in the S–N bond length $r$(S–N) at the CCSD level (1.856 Å and 1.800 Å, respectively), while improving the correlation treatment from CCSD to CCSD(T) lead to >0.04 Å lengthening (1.800 Å and 1.841 Å with quintuple-zeta basis set).

In this work, we applied explicitly–correlated (F12) coupled cluster methods instead of conventional coupled cluster approaches used earlier for HSNO modeling.[25,28] This methodological improvement largely solved the slow convergence with respect to the one-electron basis set size. Indeed, the $r$(S–N) values obtained with a double zeta VDZ-F12 basis set overestimate the corresponding CBS-extrapolated values (obtained with the same coupled-cluster method) by just ~0.003 Å. Similarly to HSNO,[25] the coupled-cluster methods with different level of electron correlation treatment demonstrate smooth and generally parallel convergence with respect to the one-electron basis set size (Figure 1A, Tables I, S2-S3).

At the same time, the calculated $r$(S–N) values demonstrate slow convergence with the excitation level included in the coupled cluster calculations; the better the correlation treatment, the longer the S–N bond (Figure 1A). At the CCSD-F12/CBS level $r$(S–N) = 1.764 Å, while at the CCSD(T)-F12/CBS level $r$(S–N) = 1.794 Å, a 0.03 Å increase; the correction to include the effect of perturbative quadruple excitations Δ(Q), estimated from limited optimization of the S–N bond in $CH_3SNO$, further lengthens the S–N bond by 0.023 Å. This behavior points out to an appreciable multi-reference character of the –SNO group earlier noted in the case of HSNO.[25,28,79] The $T_1$ and $D_1$ coupled cluster diagnostic values (Table S4) that can be used to assess the multireference character of a molecule,[80,81] are similar for $CH_3SNO$ and HSNO molecules: $T_1$=0.025 and $D_1$=0.080 for cis-$CH_3SNO$ and $T_1$=0.026 and $D_1$=0.077 for cis-HSNO,[28] above the accepted thresholds of 0.02 and 0.05, respectively, which suggests a moderate multi-reference character in both cases. A slightly smaller value of the $T_1$ diagnostic and smaller effects of the triple (+0.030 Å in cis-$CH_3SNO$ vs +0.041 Å in trans-HSNO[25]) and quadruple (+0.023 Å in cis-$CH_3SNO$ vs +0.033 Å in trans-HSNO[28]) excitations may suggest marginally reduced multi-reference character of $CH_3SNO$ compared to HSNO.

We also investigated the performance of recently developed distinguishable cluster with single and double excitations (DCSD) method by Kats et al.,[46-48] which at a similar computational cost demonstrated significant improvement over CCSD for multi-reference systems. In the case of $CH_3SNO$, DCSD-F12 indeed provides better description of the S–N bond length than CCSD-F12, 1.785 Å vs. 1.764 Å, a +0.021 Å improvement, just below (-0.009 Å) the CCSD(T)-F12 value of 1.794 Å. Conventional DCSD method shows expectedly slower convergence (Figure 1A, inset), but appears to converge to roughly the same CBS limit as DCSD-F12.

The CCSD(T)-F12/CBS+ΔQ geometries were further corrected to include core-valence correlation ΔCV and scalar-relativistic effects ΔSR. The former shortens the S–N bond in cis-$CH_3SNO$ by 0.007 Å, while the latter elongates it by 0.005 Å (-0.005 Å and 0.004 Å in trans-$CH_3SNO$). The final FPD values of *r*(S–N) in $CH_3SNO$, 1.814 Å in cis- and 1.824 Å in trans-$CH_3SNO$, are noticeably shorter our recent FPD values for HSNO, 1.842 Å in cis- and 1.858 Å in trans-HSNO, as well as recent semi-experimental values by Nava et al.,[30] 1.834(2) Å in cis- and 1.852(2) Å in trans-HSNO, derived from the experimental ground-state rotational constants corrected for zero-point vibrational motion using CCSD(T)/aug-cc-pV(Q+d)Z calculations.

The N–O bond length in RSNOs is less sensitive to the basis set size, e.g. for HSNO[28] even conventional CCSD(T) calculations with a triple-zeta basis set give a reasonable approximation to the CBS limit (1.183 Å and 1.180 Å, respectively.[28] Not surprisingly, *r*(N–O) obtained with explicitly-correlated CCSD(T)-F12 converge almost instantly to the CBS limit. In cis-$CH_3SNO$, even the smallest basis set VDZ-F12 provides an acceptable N–O bond value of 1.193 Å, just within 0.002 of the corresponding CBS limit of 1.191 Å. With respect to the correlation treatment, N–O bond elongates by 0.008 Å when going from CCSD-F12 to CCSD(T)-F12 (1.183 Å and 1.191 Å at the CBS limit, respectively), which is significantly smaller than the corresponding S–N bond elongation, 0.03 Å. At the DCSD-F12/CBS level the N–O bond length (1.190 Å) is almost identical to the CCSD(T)-F12 value. Other $CH_3SNO$ geometry parameters demonstrate even less sensitivity to the level of theory. The C–S bond length calculated at the CCSD(T)-F12/VDZ-F12 level is the same as the extrapolated CCSD(T)-F12/CBS value (1.791 Å). Both ΔCV and ΔSR corrections to *r*(N–O) do not exceed 0.002 Å in magnitude, and 0.005 Å in the case of the C–S bond. Figure 2 summarizes the final recommended FPD geometries of the cis- and trans-$CH_3SNO$ molecules obtained here.

## IV. S–N BOND DISSOCIATION ENERGY IN $CH_3SNO$

The weakness of the S–N bond which makes it prone to homolytic dissociation (many primary and secondary RSNOs have half-lives from seconds to minutes[82]) is one of the defining features of the RSNO chemistry. This makes the S–N bond homolytic dissociation energy in RSNOs one of the most important parameters, which, at the same time, is also challenging to accurately predict computationally due to slow convergence with respect to the basis set size and the degree of correlation treatment. In our earlier conventional coupled cluster studies of HSNO,[25,28] double- to quintuple-zeta basis set improvement increased $D_e$(S–N) by >5 kcal/mol (22.6 and 27.8 kcal/mol at the CCSD level). Improving the electron correlation treatment to perturbatively include the effect of triple excitations further increased $D_e$(S–N) by >5.5 kcal/mol (31.4 kcal/mol with quintuple-zeta basis set), while inclusion of perturbative quadruple excitations (evaluated with a double-zeta basis set) increased it by another ~1.3 kcal/mol.[25,28]

Explicitly-correlated coupled cluster methods expectedly improve the $D_e$(S–N) convergence with respect to the basis set size. For cis-$CH_3SNO$ the $D_e$(S–N) values obtained at the CCSD-F12 level with VDZ-F12 and VQZ-F12 differ only by 1.2 kcal/mol (29.0 vs. 27.7 kcal/mol, Figure 1B and Table S5); the convergence is even faster at the CCSD(T)-F12 level, for which the VDZ-F12 and VQZ-F12 results differ by 0.6 kcal/mol (34.7 vs. 34.1 kcal/mol). Interestingly, the $D_e$(S–N) values calculated with F12 converge to the CBS limit from above, i.e. smaller basis set calculations overestimate $D_e$(S–N). On the other hand, $D_e$(S–N) values for HSNO obtained with conventional coupled cluster calculations[25,28] converged from below, with smaller basis sets underestimating $D_e$(S–N).

The level of correlation treatment has a dramatic effect on the $D_e$(S–N) value, with CCSD(T)-F12 giving ~6.5 kcal/mol stronger S–N bond than CCSD-F12 (34.1 vs. 27.7 kcal/mol at the CBS limit, Table S5); including the effect of quadruple excitations Δ(Q) further increases $D_e$(S–N) by 1.33 kcal/mol (Tables II and S6). At the extrapolated CBS limit, DCSD-F12 preforms better than CCSD-F12, giving 4.2 kcal/mol higher $D_e$(S–N) value (31.9 vs. 27.7 kcal/mol), but still 2.2 kcal/mol below the CCSD(T)-F12/CBS value (34.1 kcal/mol). Similarly to explicitly-correlated CCSD-F12 and CCSD(T)-F12, $D_e$(S–N) calculated with DCSD-F12 converge to the CBS limit from above. However, the explicitly-correlated DCSD-F12 demonstrates much worse convergence of $D_e$(S–N) with the basis set size. Surprisingly, the conventional version of DCSD demonstrates slightly better convergence: increasing the basis set size from VDZ-F12 to VQZ-F12 decreases $D_e$(S–N) by 4.8 kcal/mol in the case of DCSD-F12, and it increases $D_e$(S–N) by 4.1 kcal/mol in the case of conventional DCSD. As in the case of other coupled cluster methods, the $D_e$(S–N) values approach to the CBS limit from below with the conventional DCSD and from above with DCSD-F12. Both methods appear to converge roughly to the same CBS value, although due to the slow basis set convergence the $CBS_{(T-Q)}$ extrapolations much less reliable than in the case of CCSD-F12 and CCSD(T)-F12 methods.

The core-valence electron correlation correction ΔCV to the $D_e$(S–N) value is minor (-0.02 kcal/mol, Table II), while the scalar-relativistic correction reduces $D_e$(S–N) by 0.41 kcal/mol. Due to the relatively high spin-orbit coupling constant of sulfur (1.13 kcal/mol)[83] and the open-shell character of $CH_3S^•$ radical, the $D_e$(S–N) value needs to be corrected for extra stabilization of the $CH_3S^•$ radical due to spin-orbit coupling, ΔSO. However, the ΔSO correction for $D_e$(S–N) is more than two-fold smaller for $CH_3SNO$ vs. the ΔSO correction reported earlier for HSNO,[25] 0.18 kcal/mol vs. 0.48 kcal/mol. This is because Jahn-Teller geometry distortion removes the degeneracy of the two lowest electronic states of $CH_3S^•$, so the energy gap between non-degenerate $^2A'$ and $^2A''$ states in $C_s$-symmetry $CH_3S^•$ is 1.54 kcal/mol (at the MRCI/aug-cc-pV(Q+d)Z level, Figure S1 in Supporting Information). On the other hand, HS· radical has two degenerate $^2\Pi$ states which leads to stronger spin-orbit coupling.

The zero-point vibrational energy (ZPE) effect on the the S–N bond dissociation energy in $CH_3SNO$ is mostly determined by the S–N stretching vibration, so the $ZPE_{harm}$ correction obtained from harmonic frequency calculations at the CCSD(T)-F12/VQZ-F12 level is similar to the HSNO case,[28] -2.79 kcal/mol and -2.77 kcal/mol, respectively. The $ZPE_{harm}$ value has been further corrected to anharmonicity, $\Delta ZPE_{anharm}$, evaluated with the second-order perturbative approach[77] using double-hybrid mPW2PLYPD/def2-TZVPPD DFT method. With this correction, out final FPD value $D_0$(S–N) for $CH_3SNO$ in the gas phase is 32.4 kcal/mol.

For a better comparison of the S–N bond strength in $CH_3SNO$ vs. HSNO, we updated the FPD value of $D_0$(S–N) reported earlier for HSNO (29.4 kcal/mol),[28] to include the anharmonicity correction $\Delta ZPE_{anharm}$ recently evaluated[38] with vibrational configuration interaction (VCI) method[84,85] at the CCSD(T)-F12/VDZ-F12 level, $\Delta ZPE_{anharm}$ = -0.29 kcal/mol, which compares very well with the second-order perturbative estimate at the mPW2PLYPD/def2-TZVPPD level, -0.28 kcal/mol. The updated FPD estimate of $D_0$(S–N) for HSNO is then 29.7 kcal/mol.

Thus, the gas-phase FPD values suggest that the S–N bond in $CH_3SNO$ is at least 2.7 kcal/mol more stable than in HSNO. We note that the correction for quadruple excitations Δ(Q) was evaluated in this work using partially optimized $CH_3SNO$ geometry (only the S–N bond was relaxed due to the computational limitations), which tends to underestimate Δ(Q) by 0.02-0.03 kcal/mol.[28] Therefore, the $D_0$(S–N) = 32.4 kcal/mol value is best considered as a lower bound for the actual value that is slightly larger (by a few tenths kcal/mol).

Finally, to make the data obtained in this work more relevant for assessing the stability of the cysteine-based biological RSNOs in the aqueous environment, we evaluated the solvation effects ΔSolv on $D_0$(S–N) using DFT calculations with polarizable continuum model (PCM). These calculations (Table S7) suggest a small decrease of $D_0$(S–N) in water (-0.17 kcal/mol) and diethylether which is often used to mimic protein environment (-0.15 kcal/mol); this suggests $D_0$(S–N) = 32.2 kcal/mol for $CH_3SNO$ in solution.

## V. CONFORMATIONAL BEHAVIOR OF $CH_3SNO$

The predisposition of RSNOs to adapt planar conformations of the –SNO fragment due to the hindered rotation around the S–N bond has been noted in numerous early experimental studies.[86-88] Cis-trans isomerism of $CH_3SNO$ was reported as early as in 1961 based on the IR spectroscopic data,[89] followed by observation of cis-trans conformational change in $CH_3SNO$ in low-temperature proton NMR experiments,[90] and IR spectroscopy in argon matrix.[91] Recent gas-phase IR studies demonstrated 3:1 cis/trans ratio for another primary RSNO, $CH_3CH_2SNO$,[92] whereas this ratio is inverted (1:4) for tertiary $(CH_3)_3CSNO$.[93]

The FPD data on the relative stability of cis-$CH_3SNO$ and trans-$CH_3SNO$ (Table II) suggest that the cis-conformer is slightly more stable, $\Delta E_0$(cis-trans) = 1.15 kcal/mol, which is typical for primary RSNOs. On the other hand, HSNO prefers trans-conformation by 0.9 kcal/mol.[28] The $\Delta E_0$(cis-trans) value is generally not sensitive to the level of theory (Table II and Table S8), and has low sensitivity to the solvent effects (trans-$CH_3SNO$ stability increases by ~0.02 kcal/mol in water and diethylether, Table S7).

We were also able to optimize and characterize the transition structure $TS_{c\text{-}t}$ of $CH_3SNO$ cis-trans interconversion with the FPD approach (Tables I and S9). The S–N bond in the $TS_{c\text{-}t}$ level is noticeably elongated compared to cis-$CH_3SNO$, by 0.13-0.16 Å, depending on the level of theory; otherwise, the evolution of $r$(S–N) and other $TS_{c\text{-}t}$ geometric parameters with increasing one-electron basis set and the level of electron correlation treatment as well as the magnitudes of the $\Delta$(Q), $\Delta$CV, and $\Delta$SR corrections are comparable to the S–N bond in the equilibrium $CH_3SNO$ structures. The final FPD $r$(S–N) value for $TS_{c\text{-}t}$, 1.980 Å, is 0.166 Å longer than in cis-$CH_3SNO$. The $TS_{c\text{-}t}$ structure is slightly non-perpendicular, with the CSNO dihedral angle 85.4°.

The activation barrier of cis-trans $CH_3SNO$ interconversion $\Delta E_e^{\neq}$ is relatively insensitive to the basis set size, and slightly increases with the level of electron correlation treatment (Table S10). The CCSD-F12/CBS predicts the $\Delta E_e^{\neq}$ of 11.8 kcal/mol, DCSD-F12/CBS gives 12.0 kcal/mol, CCSD(T)-F12/CBS gives 12.6 kcal/mol, and addition of the quadruple excitations correction $\Delta$(Q) rises $\Delta E_e^{\neq}$ to 13.2 kcal/mol; parallel to this progression, the S–N bond lengthens from 1.894 Å to 1.978 Å.

Inclusion of the $\Delta$CV and $\Delta$SR corrections (+0.1 and -0.07 kcal/mol, respectively), as well as a $ZPE_{harm}$ correction (-0.55 kcal/mol) gives the final FPD value for the cis-trans interconversion barrier $\Delta E_0^{\neq}$ = 12.65 kcal/mol.

Previous FPD investigation[28] of HSNO yielded noticeably lower value of the rotational barrier along the S–N bond, $\Delta E_0^{\neq}$ = 9.52 kcal/mol. In the case of HSNO the S–N bond elongation in the corresponding TS is larger than in the case of $CH_3SNO$ (0.175 Å vs. 0.166 Å), and the TS structure itself has a more upright geometry with the HSNO dihedral angle 88.0° vs. CSNO dihedral 85.4°.

Finally, the evaluation of the solvation effects on $CH_3SNO$ cis-trans interconversion barrier using PCM DFT approach (Table S7) suggests that polar and non-polar solvents pull the activation barrier in different directions: aqueous environment on average increases the barrier by 0.57 kcal/mol to $\Delta E_0^{\neq} = 13.22$ kcal/mol, whereas a less polar solvent (diethylether) decreases the barrier by 0.12 kcal/mol to $\Delta E_0^{\neq} = 12.63$ kcal/mol.

## VI. $CH_3SNO$ VIBRATIONAL FREQUENCIES

$CH_3SNO$ was first characterized in the gas phase by IR spectroscopy in 1961 by Philippe.[89] At the time, only a few fundamental frequencies in the IR spectrum were assigned, and the S–N and N–O bond stretches were assigned to 655 $cm^{-1}$ and 1534 $cm^{-1}$, respectively. Later, Christensen at al.[90] tentatively assigned the S–N–O bending band at 375 $cm^{-1}$, S–N stretching at 734 $cm^{-1}$, and N–O stretching at 1530 $cm^{-1}$. In 1984, Muller and Huber[91] reported the spectra of both cis-$CH_3SNO$ and trans-$CH_3SNO$ in argon matrix at 12 K, with the N–O band at 1527 $cm^{-1}$ for cis-$CH_3SNO$ and 1548 $cm^{-1}$ for trans-$CH_3SNO$ (21 $cm^{-1}$ difference), and the S–N band at 376 $cm^{-1}$ for cis-$CH_3SNO$ and 371 $cm^{-1}$ for trans-$CH_3SNO$. Recently, Cánneva et al.[92] reported gas-phase N–O stretching frequencies of 1537 $cm^{-1}$ and 1559 $cm^{-1}$ (22 $cm^{-1}$ difference) for cis- and trans-conformers of related species, $CH_3CH_2SNO$.

Here, we calculated harmonic vibrational frequencies for $CH_3SNO$ using CCSD(T)-F12/V*n*Z-F12 ($n$ = D, T, Q) with subsequent two-point $CBS_{(T-Q)}$ extrapolation (Tables III and S11-S12). The CCSD(T)-F12 harmonic vibrational frequencies of $CH_3SNO$ demonstrate fast convergence with the basis set size, with the VDZ-F12 values already near the CBS limit (the S–N bond stretch frequency in cis-$CH_3SNO$ is 400.6 cm-1 vs. 399.7 $cm^{-1}$, N–O bond stretch frequency is 1571.5 vs. 1575.1 $cm^{-1}$); and the $ZPE_{harm}$ values obtained with VDZ-F12 basis set are within 0.1 kcal/mol of the CBS limit. We further corrected the harmonic values by adding a correction for anharmonicity determined with second-order perturbative approach[77] using double-hybrid mPW2PLYPD/def2-TZVPPD DFT method (Table S13). The resulting vibrational frequencies and $ZPE_{anharm}$ values for cis-$CH_3SNO$ are listed in Table III along with available experimental data.[91]

Both experimental and FPD frequencies of the S–N bond stretching in $CH_3SNO$ are below 400 $cm^{-1}$ (398.2/376.0 $cm^{-1}$ calculated/experimental for cis-, and 386.5/371.0 $cm^{-1}$ for trans-conformer). The calculated N–O stretching frequencies, 1542 $cm^{-1}$ for cis-, and 1562 $cm^{-1}$ for trans-$CH_3SNO$, are also in reasonable agreement with the earlier experimental data on $CH_3SNO$,[91] 1527 $cm^{-1}$ and 1548 $cm^{-1}$, and in even better agreement with the recent experimental data on $CH_3CH_2SNO$, 1537 $cm^{-1}$ and 1559 $cm^{-1}$; the calculated difference in the N–O stretching frequencies for cis- and trans-$CH_3SNO$, 20 $cm^{-1}$, closely matches with the experimental values for $CH_3SNO$ (21 $cm^{-1}$) and $CH_3CH_2SNO$ (22 $cm^{-1}$).[92]

Although the S–N bond $CH_3SNO$ is relatively non-rigid, it is quite harmonic: the second-order perturbative anharmonic corrections evaluated at the mPW2PLYPD/def2-TZVPPD level (Table IV) for the S–N stretching vibration are -1.5 $cm^{-1}$ for cis- and -7.6 $cm^{-1}$ for trans-$CH_3SNO$, while the anharmonic corrections for the N–O stretch are somewhat larger, -33.2 $cm^{-1}$ and -31.8 $cm^{-1}$ for cis- and trans-$CH_3SNO$, respectively; anharmonic correction lowers the $ZPE_{harm}$ of both $CH_3SNO$ conformers by 0.9 kcal/mol.

## VII. PERFORMANCE OF DFT METHODS

We used the high-level *ab initio* data for $CH_3SNO$ generated here to evaluate the performance of common DFT methods with respect to the –SNO group properties. The DFT method performance was tested with two basis sets, a large triple-zeta basis set with two sets of polarization functions def2-TZVPPD,[72,73] and smaller double-zeta def2-SV(P)+d basis set.[72] Since routine DFT calculations typically do not include relativistic effects, we used a modified set of FPD reference data with omitted ΔSR and ΔSO corrections; the results of the DFT calculations along with the modified FPD reference data are listed in Tables IV, V and S14-S21.

Global hybrid Perdew, Burke and Ernzerhof functional PBE0[64-66] underestimates the S–N bond length in $CH_3SNO$ and in the $TS_{c\text{–}t}$ structure by 0.01-0.05 Å and the addition of an empirical dispersion term GD3[67] expectedly does not affect the geometry. The PBE0 version with the fraction of exact exchange increased from 1/4 to 1/3, PBE0-1/3, as recently proposed by Guido et al.[68] and the range-separated ωB97XD functional with empirical dispersion [69] tend to give even shorter bond lengths. While these functionals also underestimate all other bond lengths, a widely used global hybrid B3LYP functional[61,63] seems to provide an inconsistent description of the –SNO group: it overestimates the S–N bond lengths (by 0.01-0.04 Å) while underestimating the N–O bond lengths (by ~0.01 Å). Double-hybrid B2PLYP (and its dispersion-corrected variant, B2PLYPD)[70,94] method with def2-TZVPPD basis set gives equilibrium $r$(S–N) within 0.001 Å of the reference (1.811 Å B2PLYP, and 1.810 Å FPD), while mPW2PLYP/mPW2PLYPD double hybrid approach[71] underestimates equilibrium $r$(S–N) by 0.015 Å. On the other hand, mPW2PLYP/mPW2PLYPD overestimates the reference $r$(S–N) = 1.949 Å in the $TS_{c\text{–}t}$ structure by 0.01 Å, while B2PLYP/B2PLYPD overestimate it by ~0.04 Å.

PBE0 gives the best $D_0$(S–N) value (Table IV), 31.8 kcal/mol with PBE0/def2-TZVPPD basis set vs. 32.7 kcal/mol reference, the addition of empirical dispersion in PBE0-GD3 (artificially) improves the result even further, 32.5 kcal/mol, while other hybrid and double-hybrid functionals underestimate $D_0$(S–N) by 3-4 kcal/mol. DFT methods overestimate the rotational barrier for the S–N bond in $CH_3SNO$ by 0.6-2 kcal/mol relative to the reference value $\Delta E_0^{\neq}$ = 12.7 kcal/mol (Table V), with the smallest errors observed for the range-separated ωB97XD hybrid functional and the double-hybrid functionals (0.6 and ~0.9 kcal/mol, respectively, def2-TZVPPD basis set).

The harmonic vibrational frequencies of $CH_3SNO$ are on average better reproduced with the double-hybrid DFT methods (Tables S18-S21), e.g. mPW2PLYPD/def2-TZVPPD gives the best S–N (388.4 $cm^{-1}$ vs. 399.7 $cm^{-1}$ reference) and N–O (1578 $cm^{-1}$ vs. 1571 $cm^{-1}$ reference) stretching frequencies. Using smaller def2-SV(P)+d basis set leads to larger errors in computed vibrational frequencies; in particular, the N–O stretching frequency, which often used as a characteristic band in IR spectroscopy studies of RSNOs, is significantly overestimated, e.g. PBE0-GD3/def2-SV(P)+d value 1756 $cm^{-1}$ is almost 200 $cm^{-1}$ larger than the reference (1575.1 $cm^{-1}$).

Overall, DFT methods provide reasonably accurate description of the –SNO group, especially when a larger basis set used. Consistent with our earlier observations,[20,28,95] PBE0 hybrid functional generally provides a consistent description of RSNO properties; when feasible, PBE0 results can be verified by more computationally demanding double hybrid DFT calculations.

## VIII. ANTAGONISTIC NATURE OF $CH_3SNO$

The paradox of the RSNO S–N bond which is elongated, weak, and has a low stretching frequency, but, at the same time, has a sizable rotation barrier can be viewed as a consequence of the antagonistic nature of the –SNO group. In this context, antagonistic nature implies that two of the three resonance structures required to describe the –SNO group are chemical opposites of each other, or *antagonistic*.[20] These two structures, referred to as ***D*** and ***I***, imply opposite bonding patterns (double S=N bond vs. ionic/no bond) and opposite formal charges (e.g., positive versus negative charge on the sulfur atom). This simple model of the RSNO structure have been shown to have surprising explanatory and predictive power. It elegantly accounts for the extreme malleability of the S–N bond in the presence of charged or neutral Lewis acids and bases,[20,96,97] provides chemically intuitive description of subtle substituent effects in RSNOs,[19,34,98] and explains the ability of RSNOs to engage in two competing reaction modes with the same molecule.[21,95,97,99] The antagonistic paradigm also provides a useful framework for designing novel RSNO reactions[97] as well as RSNOs with desired properties.[19,34,98]

The accurate FPD data on the $CH_3SNO$ structure reported here are consistent with the antagonistic model. Compared to HSNO, the S–N bond is ~0.03 Å shorter (1.814 Å vs 1.842 Å in cis-$CH_3SNO$ and cis-HSNO, respectively) and 2.7 kcal/mol stronger ($D_0$ is 32.4 vs 29.7 in cis-$CH_3SNO$ and trans-HSNO), and the rotation barrier is ~3 kcal/mol higher ($E_0^{\neq}$ is 12.7 vs. 9.5 kcal/mol).[28] This is consistent with electron-donating character of the $CH_3$– group that favors the structure ***D*** with a positive formal charge on the sulfur atom and double S=N bond. The transition structure for the rotation along the S–N bond correlates well with the removal of the resonance structure ***D*** that otherwise counteracts the effect of the ionic no-bond resonance structure ***I***. This leads to dramatic lengthening of the S–N bond to >1.9 Å, well beyond the distance expected for a covalent bond involving these atoms. Importantly, the variation of the S–N bond in cis- and trans-$CH_3SNO$ and the $TS_{c\text{-}t}$ structure anti-correlates with the N–O bond length, in agreement with the antagonistic resonance description (Figure 3A).

However, a more direct way to probe the antagonistic nature of the –SNO group is to observe the effect of an external electric field (EEF) on its properties. If the opposite formal charges implied by the antagonistic structures are to be given credence, one should expect significant change in the contribution of these structures with attendant significant changes in the S–N bond length. Indeed, optimization of the $CH_3SNO$ geometry in an EEF oriented along the S–N bond, $F_Z$, varied from +0.015 to -0.015 au (1 au = 51.4 V/Å), leads to dramatic changes in the S–N bond length, much larger than the corresponding changes observed for typical single and double S–N bonds (Figure 3B). The effect is particularly well pronounced for the negative $F_Z$ values that lead to >0.2 Å lengthening of the S–N bond due to the increasing contribution of the structure ***I*** and decreasing contribution of the structure ***D***. On the other hand, shortening of the S–N bond in the positive fields is smaller, up to 0.1 Å. Scanning the EEF oriented along the S–O axis gives essentially the same results (Figure S2).

DFT methods reproduce the S–N bond variation $\Delta r$(S–N) in EEF, with PBE0 only slightly underestimating the shortening of the positive fields, and the double hybrid mPW2PLYP method slightly overestimating $\Delta r$(S–N) across the board (Figure 3B). The evolution of the –SNO group electronic structure can be conveniently quantified by the analysis of the DFT density matrix with natural resonance theory (NRT), which expresses the density matrix in terms of the resonance contributions of several Lewis structures. NRT calculations (Figure 3C) show that the dominant structure ***S*** has a fairly constant contribution %***S*** within the $F_Z$ range studied, whereas the ***D*** structure contribution changes linearly with the electric field. The variation of the structure ***I*** contribution %***I*** is slightly non-linear, mirroring the nonlinearity in the %***S*** evolution, with a positive (%***I***) and a negative (%***S***) curvatures. This nonlinearity correlates with a similar curving of $\Delta r$(S–N) vs. $F_Z$ dependence, and can be attributed to the slower decrease of %***I*** in the $F_Z > 0$, so the linear increase in %***D*** has to be compensated by additional %***S*** decrease.

At $F_Z = 0$ structure ***D*** has a higher contribution than ***I***; as $F_Z$ increases in the negative direction, %***I*** increases at the expense of %***D***. At $F_Z \approx -0.0085$ au the ***D*** and ***I*** contributions balance out, and ***I*** becomes the dominant antagonistic structure beyond that point. Remarkably, this changeover in %***D*** and %***I*** nearly coincides with the molecular dipole moment projection (Figure 3D) onto the S–N axis $\mu_Z$ reaching zero at -0.007 au (-0.0065 au for CCSD(T)–F12). For the fields above this critical value, the orientation of the dipole moment is consistent with positively charged sulfur atom and negatively charged NO moiety, i.e. the predominance of the structure ***D***; in the more negative fields the dipole moment is reversed, consistent with the predominance of the structure ***I***.

We also used EEF to probe the transition structure for rotation along the S–N bond $TS_{c\text{-}t}$ (Figure 4A). Although the S–N bond in the $TS_{c\text{-}t}$ is significantly longer, the evolution of its relative change $\Delta r$(S–N) is very similar to that of cis-$CH_3SNO$, with the main difference that the S–N lengthening in the TS is slower for $F_Z < -0.005$ au (Figure 4B). Although the contribution of structure ***D*** is nearly negligible (but not zero due to a slightly non-perpendicular dihedral angle, ~85°, Figure 4C), %***I*** is only slightly larger (by ≈3%) than in cis-$CH_3SNO$; however, in the absence of structure ***D***, it is sufficient to significantly weaken the S–N bond. The evolution of the dipole moment projection $\mu_Z$ is similar to cis-$CH_3SNO$, but shifted by approximately 0.007 a.u. toward the positive values, reflecting the stronger effect of the structure ***I*** (Figure 4D).

The EEF effect on the S–N bond in cis-$CH_3SNO$ is determined by an interplay between the agonistic structures ***D*** and ***I***, but for $TS_{c\text{-}t}$ the EEF effect is mainly due an interplay between the one remaining antagonistic structure ***I*** and the dominant conventional structure ***S*** (Figures 3A and 4A, correspondingly). Although this underlying difference is not immediately evident from the evolution of the S–N bond length, it can be gleaned from the evolution of the N–O bond length (Figure 5). Indeed, $r$(N-O) vs. $F_Z$ has a larger slope for cis-$CH_3SNO$ because ***D*** and ***I*** interconversion causes more significant change in the N–O bond nature (single vs. triple). The slope is smaller in the case of the $TS_{c\text{-}t}$ structure, since the N–O bond nature change is less dramatic for ***I*** and ***S*** interconversion (triple vs. double).

Thus, analysis of the physical observables (geometry and dipole moment) obtained with *ab initio* and DFT calculations supports the antagonistic model of the –SNO group, a powerful conceptual model that can be conveniently quantified using the NRT analysis. This study also shows that DFT methods are capable of correctly capturing the evolution of the –SNO group properties across a wide range of external perturbations. Finally, it has been hypothesized that biochemical reactions can be controlled by electric fields created in proteins, which can reach up to ±0.01 au.[100-106] This suggests a possible mechanism of effective biological control of protein CysNO reactivity that takes advantage of the peculiar antagonistic nature of the –SNO group.

## IX. CONCLUSIONS

In this work, we reported accurate *ab initio* calculations of the structure and properties of the $CH_3SNO$ (summarized in Figure 2) using Feller-Peterson-Dixon (FPD) approach based on the explicitly-correlated coupled-cluster methodology with extrapolation to the complete basis set limit with several additive corrections. These accurate computational data on $CH_3SNO$, the smallest aliphatic S-nitrosothiol (RSNO), provide a useful estimation of the intrinsic properties of the S–N bond in S-nitrosated cysteine aminoacid residue (CysNO) sidechain. Compared to a smaller RSNO model molecule—and likely also a biological RSNO itself—thionitrous acid HSNO, the S–N bond in $CH_3SNO$ is ~0.03 Å shorter, ~3 kcal/mol stronger, and has ~3 kcal/mol higher rotational barrier. While the energetic difference between cis- and trans-conformers is roughly the same, ~1 kcal/mol, $CH_3SNO$ prefers the cis-orientation of the NO moiety, whereas HSNO prefers the trans-form.

While introduction of efficient explicitly-correlated coupled-cluster methods alleviates the slow convergence of the S–N bond properties with the one-electron basis set size, slow convergence with the coupled-cluster excitation level remains a problem. While the recently developed distinguished cluster approximation, DCSD, works significantly better than the traditional CCSD method, it falls short of the coupled-cluster methods that include triple and quadruple excitations. Fortunately, some commonly used density functional theory methods, such as PBE0 and mPW2PLYPD tested in this work, provide a reasonably accurate description of the –SNO group at a modest computational cost.

Curiously, the evolution of the S–N bond properties with respect to the level of correlation treatment is rather counterintuitive. As the coupled-cluster description improves, this bond becomes longer and floppier and, at the same time, harder to dissociate or rotate around. On transition from CCSD to CCSDT(Q), the calculated S–N bond in $CH_3SNO$ becomes >0.05 Å longer (1.794 Å to 1.817 Å, estimated CBS limit, see also Table S2) and its stretching force constant drops by >0.2 mdyn/Å (0.8 to 0.56 mdyn/Å, estimated with a double-zeta basis set, Table S22), while its bond dissociation energy increases by >7.7 kcal/mol ($D_e$ = 27.7 to 35.4 kcal/mol, estimated CBS limit) and the rotation barrier increases by ~1.4 kcal/mol (11.8 to 13.2 kcal/mol, estimated CBS limit). All this points out to a rather unusual and complex, multi-reference character of the –SNO group.

Conceptually, the properties of $CH_3SNO$ (as well as other RSNO molecules) can be understood through the antagonistic resonance model that represents its chemical structure as a hybrid of three Lewis structures (Figure 3A), two of which are chemical opposites of each other—antagonistic structures. This model can be quantified using the natural resonance theory (NRT) approach, and tested by monitoring the –SNO group response to the external electric fields (EEFs). Remarkably, the inversion of the dipole moment projection observed for $CH_3SNO$ in a moderately strong EEF seems to correlate with inversion of the relative order of the resonance contributions of the two antagonistic structures calculated with NRT.

It is interesting to consider if there is a relation between the antagonistic nature and the multireference character of the –SNO group. As we have seen in this work, the calculated S–N bond length in the cis-trans interconversion transition structure $TS_{c\text{-}t}$ converges with the coupled-cluster excitation level at least as slowly as in the equilibrium structures (e.g., Table S22). As the S–N bond in $TS_{c\text{-}t}$ loses its double-bond character, this seems to support our hypothesis[79] that connects the ionic component $RS^-$ / $NO^+$ and the multi-reference character of the –SNO group. Since the unusual electronic structure of the –SNO group likely plays a defining role in the biological reactivity of RSNOs, further investigations in this direction are warranted.

## SUPPLEMENTARY MATERIAL

See supplementary material for the *ab initio* and DFT structural, spectroscopic and energetic parameters of the $CH_3SNO$ isomers (Tables S1-S22, Figures S1-S2).

## ACKNOWLEDGMENTS

This work has been supported by the National Science Foundation (NSF) CAREER award CHE-1255641, and the Extreme Science and Engineering Discovery Environment (XSEDE) allocation under 'Computational Modeling of Biologically Important S-Nitrosothiol Reactions' project TG-CHE140079 (Q.K.T.). Calculations were performed on the high-performance dedicated XSEDE cluster *Comet* and on the computational cluster *Pere̓* at Marquette University funded by NSF awards OCI-0923037 and CBET-0521602.

## BIBLIOGRAPHY

[1]K. A. Broniowska, A. R. Diers and N. Hogg, Biochim. Et Biophys. Acta 1830, 3173 (2013).

[2]K. A. Broniowska and N. Hogg, Antiox. & Redox Signal. 17, 969 (2012).

[3]J. R. Lancaster, Arch. Biochem. Biophys. 617, 137 (2016).

[4]B. C. Smith and M. A. Marletta, Curr. Opin. Chem. Biol. 16, 498 (2012).

[5]T. Nakamura and S. A. Lipton, Trends Pharmacol. Sci. 37, 73 (2016).

[6]D. T. Hess, A. Matsumoto, S.-O. Kim, H. E. Marshall and J. S. Stamler, Nat. Rev. Mol. Cell Biol. 6, 150 (2005).

[7]S. M. Haldar and J. S. Stamler, J. Clin. Invest. 123, 101(2013).

[8]D. Seth and J. S. Stamler, Curr. Opin. Chem. Biol. 15, 129 (2011).

[9]T. Nakamura and S. A. Lipton, Antioxid. Redox. Signal. 18, 239 (2013).

[10]N. Gould, P.-T. Doulias, M. Tenopoulou, K. Raju and H. Ischiropoulos, J. Biol. Chem. 288, 26473 (2013).

[11]A. Feechan, E. Kwon, B.-W. Yun, Y. Wang, J. A. Pallas and G. J. Loake, Proc. Natl. Acad. Sci. USA 102, 8054 (2005).

[12]M. Zaffagnini, M. De Mia, S. Morisse, N. Di Giacinto, C. H. Marchand, A. Maes, S. D. Lemaire and P. Trost, Biochim. Biophys. Acta 1864, 952 (2016).

[13]D. Seth, A. Hausladen, Y.-J. Wang and J. S. Stamler, Science 336, 470 (2012).

[14]P. Anand and J. S. Stamler, J. Mol. Med. 90, 233 (2012).

[15]J. S. Stamler and D. T. Hess, Nat. Cell Biol. 12, 1024 (2010).

[16]D. L. H. Williams, J. Chem. Soc., Chem. Comm. 1758 (1993).

[17]D. L. H. Williams, Acc. Chem. Res. 32, 869 (1999).

[18]M. D. Bartberger, J. D. Mannion, S. C. Powell, J. S. Stamler, K. N. Houk and E. J. Toone, J. Am. Chem. Soc. 123, 8868 (2001).

[19]C C. Gaucher, A. Boudier, F. Dahboul, M. Parent and P. Leroy, Curr. Pharm. Des. 19, 458 (2013).

[20]M. R Talipov and Q. K. Timerghazin, J. Phys. Chem. B 117, 1827 (2013).

[21]Q. K. Timerghazin and M. R Talipov, J. Phys. Chem. Lett. 4, 1034 (2013).

[22]W. H. Koppenol, Inorg. Chem. 51, 5637(2012).

[23]Y. Fu, Y. Mou, B.-L. Lin, L.Liu and Q.-X. Guo, J. Phys. Chem. A 106, 12386 (2002).

[24]C. Baciu and J. W. Gauld, J. Phys. Chem. A 107, 9946 (2003).

[25]Q. K. Timerghazin, G. H Peslherbe and A. M. English, Phys. Chem. Chem. Phys. 10, 1532 (2008).

[26]B. Nagy, P. Szakács, J. Csontos, Z. Rolik, G. Tasi and M. Kállay, J. Phys. Chem. A 115, 7823 (2011).

[27]M. Hochlaf, R. Linguerri and J. S. Francisco, J. Chem. Phys. 139, 234304 (2013).

[28]L. V. Ivanova, B. J. Anton and Q. K. Timerghazin, Phys. Chem. Chem. Phys. 16, 8476 (2014).

[29]M. Méndez, J. S. Francisco and D. A. Dixon, Chem. Eur. J. 20, 10231 (2014).

[30]M. Nava, M.-A. Martin-Drumel, C. A. Lopez, K. N. Crabtree, C. C. Womack, T. L. Nguyen, S. Thorwirth, C. C. Cummins, J. F. Stanton and M. C. McCarthy, J. Am. Chem. Soc. 138, 11441 (2016).

[31]J. L. Miljkovic, I. Kenkel, I. Ivanović-Burmazović and M. R. Filipovic, Angew. Chem. Int. Ed. 52, 12061 (2013).

[32]M. R. Filipovic, J. L. Miljkovic, T. Nauser, M. Royzen, K. Klos, T. Shubina, W. H. Koppenol, S. J. Lippard and I. Ivanović-Burmazović, J. Am. Chem. Soc. 134, 12016 (2012).

[33]B. S. King, Free Rad. Biol. Med. 55, 1 (2012).

[34]M. Flister and Q. K. Timerghazin, J. Phys. Chem. A 118, 9914 (2014).

[35]T. B. Adler, G. Knizia and H.-J. Werner, J. Chem. Phys. 127, 221106 (2007).

[36]G. Knizia, T. B. Adler and H.-J. Werner, J. Chem. Phys. 130, 054104 (2009).

[37]G. Rauhut, G. Knizia and H.-J. Werner, J. Chem. Phys. 130, 054105 (2009).

[38]D. Khomyakov, M.Sc. thesis, Marquette University, 2015.

[39]D. Feller, K. A. Peterson and D. A. Dixon, Mol. Phys. 110, 2381 (2012).

[40]K. A. Peterson, D. Feller and D. A. Dixon, Theor. Chem. Acc. 131, 1 (2012).

[41]D. A. Dixon, D. Feller, K. A. Peterson, R. A. Wheeler and G. S. Tschumper, Ann. Rep. Comp. Chem. 8, 1 (2012).

[42]H.-J. Werner, P. J. Knowles, G. Knizia, F. R. Manby and M. Schütz, WIREs Comput. Mol. Sci. 2, 242 (2012).

[43]J. F. Stanton, J. Gauss, M. E. Harding, P. G. Szalay, A. A. Auer, R. J. Bartlett, U. Benedikt, C. Berger, D. E. Bernholdt and Y. J. Bomble, CFOUR 1.0, 2009, http://www.cfour.de.

[44]M. K´allay, MRCC, a String-based Quantum Chemical Program Suite, http://www.mrcc.hu.

[45]K. A. Peterson, T. B. Adler and H.-J. Werner, J. Chem. Phys. 128, 084102 (2008).

[46]D. Kats and F. R. Manby, J. Chem. Phys. 139, 021102 (2013).

[47]D. Kats, J. Chem. Phys. 141, 061101 (2014).

[48]D. Kats, D. Kreplin, H.-J. Werner and F. R. Manby, J. Chem. Phys. 142, 064111 (2015).

[49]K. A Peterson, D. E. Woon and T. H. Dunning Jr, J. Chem. Phys. 100, 7410 (1994).

[50]J. G. Hill, S. Mazumder and K. A. Peterson, J. Chem. Phys. 132, 054108 (2010).

[51]M. Douglas and N. M. Kroll, Ann. Phys. 82, 89 (1974).

[52]G. Jansen and B. A. Heß, Phys. Rev.A 39, 6016 (1989).

[53]D. W. Schwenke, J. Chem. Phys. 122, 14107 (2005).

[54]J. G. Hill, K. A. Peterson, G. Knizia and H.-J. Werner, J. Chem. Phys. 131, 194105 (2009).

[55]D. G. Fedorov, S. Koseki, M. W. Schmidt and M. S. Gordon, Int. Rev. Phys. Chem. 22, 551 (2003).

[56]H.-J. Werner and P. J. Knowles, J. Chem. Phys. 89, 5803 (1988).

[57]P. J. Knowles and H.-J. Werner, Chem. Phys. Lett. 145, 514 (1988).

[58]T. H. Dunning Jr, J. Chem. Phys. 90, 1007 (1989).

[59]T. H. Dunning Jr, K. A. Peterson and A. K. Wilson, J. Chem. Phys. 114, 9244 (2001).

[60]M. J. Frisch, G. W. Trucks, H. B. Schlegel, G. E. Scuseria, M. A. Robb, J. R. Cheeseman, G. Scalmani, V. Barone, B. Mennucci, G. A. Petersson, H. Nakatsuji, M. Caricato, X. Li, H. P. Hratchian, A. F. Izmaylov, J. Bloino, G. Zheng, J. L. Sonnenberg, M. Hada, M. Ehara, K. Toyota, R. Fukuda, J. Hasegawa, M. Ishida, T. Nakajima, Y. Honda, O. Kitao, H. Nakai, T. Vreven, J. A. Montgomery, Jr., J. E. Peralta, F. Ogliaro, M. Bearpark, J. J. Heyd, E. Brothers, K. N. Kudin, V. N. Staroverov, R. Kobayashi, J. Normand, K. Raghavachari, A. Rendell, J. C. Burant, S. S. Iyengar, J. Tomasi, M. Cossi, N. Rega, J. M. Millam, M. Klene, J. E. Knox, J. B. Cross, V. Bakken, C. Adamo, J. Jaramillo, R. Gomperts, R. E. Stratmann, O. Yazyev, A. J. Austin, R. Cammi, C. Pomelli, J. W. Ochterski, R. L. Martin, K. Morokuma, V. G. Zakrzewski, G. A. Voth, P. Salvador, J. J. Dannenberg, S. Dapprich, A. D. Daniels, Ö. Farkas, J. B. Foresman, J. V. Ortiz, J. Cioslowski, and D. J. Fox, Gaussian 09, Gaussian, Inc., Wallingford CT, 2009.

[61]A. D. Becke, Phys. Rev. A 38, 3098 (1988).

[62]C. Lee, W. Yang and R. G. Parr, Phys. Rev. B 37, 785 (1988).

[63]P. J. Stephens, F. J. Devlin, C. F. Chabalowski and M. J. Frisch, J. Phys. Chem. 98, 11623 (1994).

[64]C. Adamo and V. Barone, J. of Chem. Phys. 110, 6158 (1999).

[65]J. P. Perdew, K. Burke and M. Ernzerhof, Phys. Rev. Lett. 77, 3865 (1996).

[66]M. Ernzerhof and G. E. Scuseria, J. Chem. Phys. 110, 5029 (1999).

[67]S. Grimme, J. Antony, S. Ehrlich and H. Krieg, J. Chem. Phys. 132, 154104 (2010).

[68]C. A. Guido, E. Brémond, C. Adamo and P. Cortona, J. Chem. Phys. 138, 021104 (2013).

[69]J.-D. Chai M. Head-Gordon, Phys. Chem. Chem. Phys. 10, 6615 (2008).

[70]S. Grimme, J. Chem. Phys. 124, 034108 (2006).

[71]T. Schwabe and S. Grimme, Phys. Chem. Chem. Phys. 8, 4398 (2006).

[72]F. Weigend and R. Ahlrichs, Phys. Chem. Chem. Phys. 7, 3297 (2005).

[73]D. Rappoport and F. Furche, J. Chem. Phys. 133, 134105 (2010).

[74]E. D. Glendening and F. Weinhold, J. Comp. Chem. 19, 593 (1998).

[75]F. Weinhold and C. R Landis, *Valency and Bonding: A Natural Bond Orbital Donor-acceptor Perspective* (Cambridge University Press, 2005).

[76]E. D. Glendening, J. K. Badenhoop, A. E. Reed, J. E. Carpenter, J. A. Bohmann, C. M. Morales and F. Weinhold, NBO 5.9, Theoretical Chemistry Institute, University of Wisconsin, Madison, 2004.

[77]V. Barone, J. Chem. Phys. 122, 14108 (2005).

[78]J. Tomasi, B. Mennucci and R. Cammi, Chem. Rev. 105, 2999 (2005).

[79]Q. K. Timerghazin, A. M. English and G. H. Peslherbe, Chem. Phys. Lett. 454, 24 (2008).

[80]T. J. Lee and P. R. Taylor, Int. J. Quant. Chem. 36, 199 (1989).

[81]C. L. Janssen and I. M. Nielsen, Chem. Phys. Lett. 290, 423 (1998).

[82]B. Roy, A. D. M. d'Hardemare and M. Fontecave, J. Org. Chem. 59, 7019 (1994).

[83]J. E. Sansonetti and W. C. Martin, J. Phys. Chem. Ref. Data 34, 1559 (2005).

[84]T. Hrenar, H.-J. Werner and G. Rauhut, J. Chem. Phys. 126, 134108 (2007).

[85]M. Neff and G. Rauhut, J. Chem. Phys. 131, 124129 (2009).

[86]L. Field, R. V. Dilts, R. Ravichandran, P. G. Lenhert and G. E. Carnahan, J. Chem. Soc., Chem. Comm. 249 (1978).

[87]G. E. Carnahan, P. G. Lenhert and R. Ravichandran, Acta Cryst. Sec. B 34, 2645 (1978).

[88]N. Arulsamy, D. S. Bohle, J. A. Butt, G. J. Irvine, P. A. Jordan and E. Sagan, J. Am. Chem. Soc. 121, 7115 (1999).

[89]R. J. Philippe, J. Mol. Spectr. 6, 492 (1961).

[90]D. H. Christensen, N. Rastrup-Andersen, D. Jones, P. Klabof and E. R. Lippincott, Spectrochim. Acta Part A: Mol. Spect. 24, 1581 (1968).

[91]R. P. Müller and J. R. Huber, J. Phys. Chem. 88, 1605 (1984).

[92]A. Cánneva, C. O. Della Védova, N. W. Mitzel and M. F. Erben, J. Phys. Chem. A 119, 1524 (2014).

[93]A. Canneva, M. F. Erben, R. M. Romano, Yu. V. Vishnevskiy, C. G. Reuter, N. W. Mitzel and C. O. Della Védova, Chem. Eur. J 21, 10436 (2015).

[94]T. Schwabe and S. Grimme, Phys. Chem. Chem. Phys. 9, 3397 (2007).

[95]L. V. Ivanova, D. Cibich, G. Deye, M. R. Talipov and Q. K. Timerghazin, ChemBioChem 18, 726 (2017).

[96]Q. K. Timerghazin, G. H. Peslherbe and A. M. English, Org. Lett. 9, 3049 (2007).

[97]M. R. Talipov, D. G. Khomyakov, M. Xian and Q. K. Timerghazin, J. Comp. Chem. 34, 1527 (2013).

[98]B. Meyer, A. Genoni, A. Boudier, P. Leroy and M. Ruiz-Lopez, J. Phys. Chem. A 120, 4191 (2016).

[99]E. E. Moran, Q. K. Timerghazin, E. Kwong and A. M. English, J. Phys. Chem. B 115, 3112 (2011).

[100]R. Meir, H. Chen, W. Lai and S. Shaik, Chem. Phys. Phys. Chem. 11, 301 (2010).

[101]P. M. de Biase, F. Doctorovich, D. H. Murgida and D. A. Estrin, Chem. Phys. Lett. 434, 121 (2007).

[102]P. M. de Biase, D. A. Paggi, F. Doctorovich, P. Hildebrandt, D. A. Estrin, D. H. Murgida and M. A. Marti, J. Am. Chem. Soc. 131, 16248 (2009).

[103]H. Hirao, H. Chen, M. A. Carvajal, Y. Wang and S. Shaik, J. Am. Chem. Soci. 130, 3319 (2008).

[104]W. Lai, H. Chen, K.-B. Cho and S. Shaik, J. Phys. Chem. Lett. 1, 2082 (2010).

[105]S. Shaik, S. P. de Visser and D. Kumar, J. Am. Chem. Soc. 126, 11746 (2004).

[106]P. J Aittala, O. Cramariuc and T. I. Hukka, J. Chem. Theor. Comp. 6, 805 (2010).

TABLE I. FPD geometric parameters of $CH_3SNO$ conformers

| Parameter | CCSD(T)-F12/$CBS_{(T-Q)}$ | Δ(Q) | ΔCV | ΔSR | Final FPD value |
|---|---|---|---|---|---|
| | | cis-$CH_3SNO$ | | | |
| $r$(S–N), Å | 1.794 | 0.023 | -0.007 | 0.005 | 1.814 |
| $r$(N–O), Å | 1.191 | - | -0.002 | -0.001 | 1.189 |
| $r$(C–S), Å | 1.791 | - | -0.004 | 0.002 | 1.789 |
| ∠SNO, ° | 117.42 | - | 0.08 | -0.06 | 117.45 |
| ∠CSN, ° | 102.33 | - | 0.10 | -0.12 | 102.31 |
| | | trans-$CH_3SNO$ | | | |
| $r$(S–N), Å | 1.799 | 0.026 | -0.005 | 0.004 | 1.824 |
| $r$(N–O), Å | 1.188 | - | -0.002 | -0.001 | 1.186 |
| $r$(C–S), Å | 1.797 | - | -0.005 | 0.002 | 1.795 |
| ∠SNO, ° | 115.62 | - | 0.001 | -0.06 | 115.56 |
| ∠CSN, ° | 94.98 | - | 0.10 | -0.10 | 94.98 |
| | | cis-trans isomerizaiton TS | | | |
| $r$(S–N), Å | 1.955 | 0.026 | -0.007 | 0.006 | 1.980 |
| $r$(N–O), Å | 1.167 | - | -0.002 | -0.001 | 1.165 |
| $r$(C–S), Å | 1.809 | - | -0.004 | 0.002 | 1.806 |
| ∠SNO, ° | 113.07 | - | 0.02 | -0.03 | 113.06 |
| ∠CSN, ° | 90.85 | - | 0.10 | -0.18 | 90.77 |
| ∠CSNO, ° | 85.39 | - | -0.01 | 0.06 | 85.44 |

TABLE II. FPD energy properties of $CH_3SNO$ conformers, kcal/mol

| FPD Component | $D$(S–N) | $\Delta E_{cis\text{-}trans}$ | $\Delta E^{\neq}_{cis\text{-}trans}$ |
|---|---|---|---|
| CCSD(T)-F12/$CBS_{(T-Q)}$ | 34.13 | 1.15 | 12.61 |
| Δ(Q) | 1.33 | -0.06 | 0.56 |
| ΔCV | -0.02 | 0.01 | 0.10 |
| ΔSO | -0.18 | - | - |
| ΔSR | -0.41 | 0.00 | -0.07 |
| $ZPE_{harm}$ | -2.79 | 0.04 | -0.55 |
| $\Delta ZPE_{anharm}$ | 0.38 | 0.01 | - |
| Final FPD value | 32.40 | 1.15 | 12.65 |

TABLE III. Cis-$CH_3SNO$ and trans-$CH_3SNO$ FPD vibrational frequencies

| Mode | CCSD(T)-F12/ $CBS_{(T-Q)}$ | ΔAnh. | Final FPD value (vs. Ref.[91]) | CCSD(T)-F12/ $CBS_{(T-Q)}$ | ΔAnh. | Final FPD value (vs. Ref.[91]) |
|---|---|---|---|---|---|---|
| | Cis-$CH_3SNO$ | | | Trans-$CH_3SNO$ | | |
| 1 A" | 84.6 | -39.6 | 45.1 | 119.5 | -17.3 | 102.2 |
| 2 A' | 280.1 | -10.0 | 270.1 (268.0) | 227.5 | -6.9 | 220.6 (234.5) |
| 3 A" | 291.0 | -4.8 | 286.2 | 235.6 | -11.1 | 224.5 |
| 4 A' (S–N) | 399.7 | -1.5 | 398.2 (376.0) | 394.2 | -7.6 | 386.5 (371.0) |
| 5 A' | 662.4 | -9.5 | 652.9 (649.0) | 669.6 | -9.5 | 660.1 (651.0) |
| 6 A' | 754.3 | -16.4 | 737.9 (731.5) | 751.8 | -14.7 | 737.1 (736.5) |
| 7 A' | 962.5 | -22.8 | 939.7 | 1011.0 | -16.3 | 994.7 |
| 8 A" | 970.8 | -12.8 | 958.0 (940.0) | 953.5 | -17.8 | 935.7 (971.5) |
| 9 A' | 1334.3 | -35.2 | 1299.1 (1298.0) | 1352.6 | -32.9 | 1319.7 (1314.0) |
| 10 A" | 1477.2 | -41.3 | 1436.0 (1428.5) | 1468.3 | -41.7 | 1426.6 (1441.0) |
| 11 A' | 1480.8 | -44.7 | 1436.1 (1455.0) | 1494.5 | -42.4 | 1452.1 (1456.0) |
| 12 A' (N–O) | 1575.1 | -33.2 | 1541.9 (1527.0) | 1593.5 | -31.8 | 1561.8 (1548.0) |
| 13 A' | 3028.4 | -96.8 | 2931.6 (2910.0) | 3046.2 | -98.2 | 2948.0 (2909.0) |
| 14 A' | 3127.3 | -137.3 | 2990.0 (2928.0) | 3142.3 | -138.9 | 3003.4 (2928.5) |
| 15 A" | 3157.2 | -145.3 | 3011.9 (2932.0) | 3150.6 | -144.2 | 3006.4 (2931.5) |
| ZPE, kcal/mol | 28.0 | -0.4 | 27.1 | 28.0 | -0.4 | 27.1 |

TABLE IV. Performance of the DFT methods (with def2-TZVPPD basis set) vs. truncated FPD scheme for cis-$CH_3SNO$

| Method | *r*(S–N), Å | *r*(N–O), Å | *r*(C–S), Å | ∠SNO, ° | $D_0$(S–N), kcal/mol |
|---|---|---|---|---|---|
| CCSD(T)-F12/ $CBS_{(T-Q)}$+ Δ(Q)+ΔCV+$ZPE_{harm}$ | 1.810 | 1.189 | 1.787 | 117.51 | 32.67 |
| B3LYP | 1.816 | 1.182 | 1.800 | 117.83 | 28.92 |
| PBE0 | 1.779 | 1.179 | 1.782 | 117.91 | 31.83 |
| PBE0-GD3 | 1.779 | 1.179 | 1.783 | 117.98 | 32.54 |
| PBE0-1/3 | 1.760 | 1.176 | 1.778 | 118.14 | 28.86 |
| ωB97XD | 1.767 | 1.181 | 1.789 | 118.12 | 28.94 |
| B2PLYP | 1.811 | 1.190 | 1.794 | 117.61 | 29.21 |
| B2PLYPD | 1.812 | 1.190 | 1.796 | 117.72 | 29.92 |
| mPW2PLYP | 1.794 | 1.188 | 1.792 | 117.79 | 28.32 |
| mPW2PLYPD | 1.795 | 1.188 | 1.794 | 117.87 | 28.83 |

TABLE V. Performance of the DFT methods (with def2-TZVPPD basis set) vs. truncated FPD scheme for $CH_3SNO$ $TS_{c\text{-}t}$

| Method | *r*(S–N), Å | *r*(N–O), Å | *r*(C–S), Å | ∠SNO, ° | ∠CSNO, ° | $\Delta E_0^{\neq}$, kcal/mol |
|---|---|---|---|---|---|---|
| CCSD(T)-F12/ $CBS_{(T-Q)}$+ Δ(Q)+ΔCV+$ZPE_{harm}$ | 1.949 | 1.166 | 1.805 | 113.09 | 85.42 | 12.71 |
| B3LYP | 1.987 | 1.155 | 1.818 | 113.84 | 85.39 | 13.70 |
| PBE0 | 1.939 | 1.154 | 1.800 | 113.71 | 85.08 | 14.62 |
| PBE0-GD3 | 1.939 | 1.153 | 1.801 | 113.72 | 85.28 | 14.66 |
| PBE0-1/3 | 1.912 | 1.152 | 1.795 | 113.67 | 85.00 | 14.22 |
| ωB97XD | 1.924 | 1.157 | 1.805 | 113.51 | 85.70 | 13.35 |
| B2PLYP | 1.984 | 1.163 | 1.814 | 113.63 | 85.04 | 13.65 |
| B2PLYPD | 1.985 | 1.162 | 1.816 | 113.65 | 85.34 | 13.71 |
| mPW2PLYP | 1.960 | 1.161 | 1.811 | 113.62 | 85.06 | 13.56 |
| mPW2PLYPD | 1.961 | 1.161 | 1.812 | 113.62 | 85.30 | 13.60 |

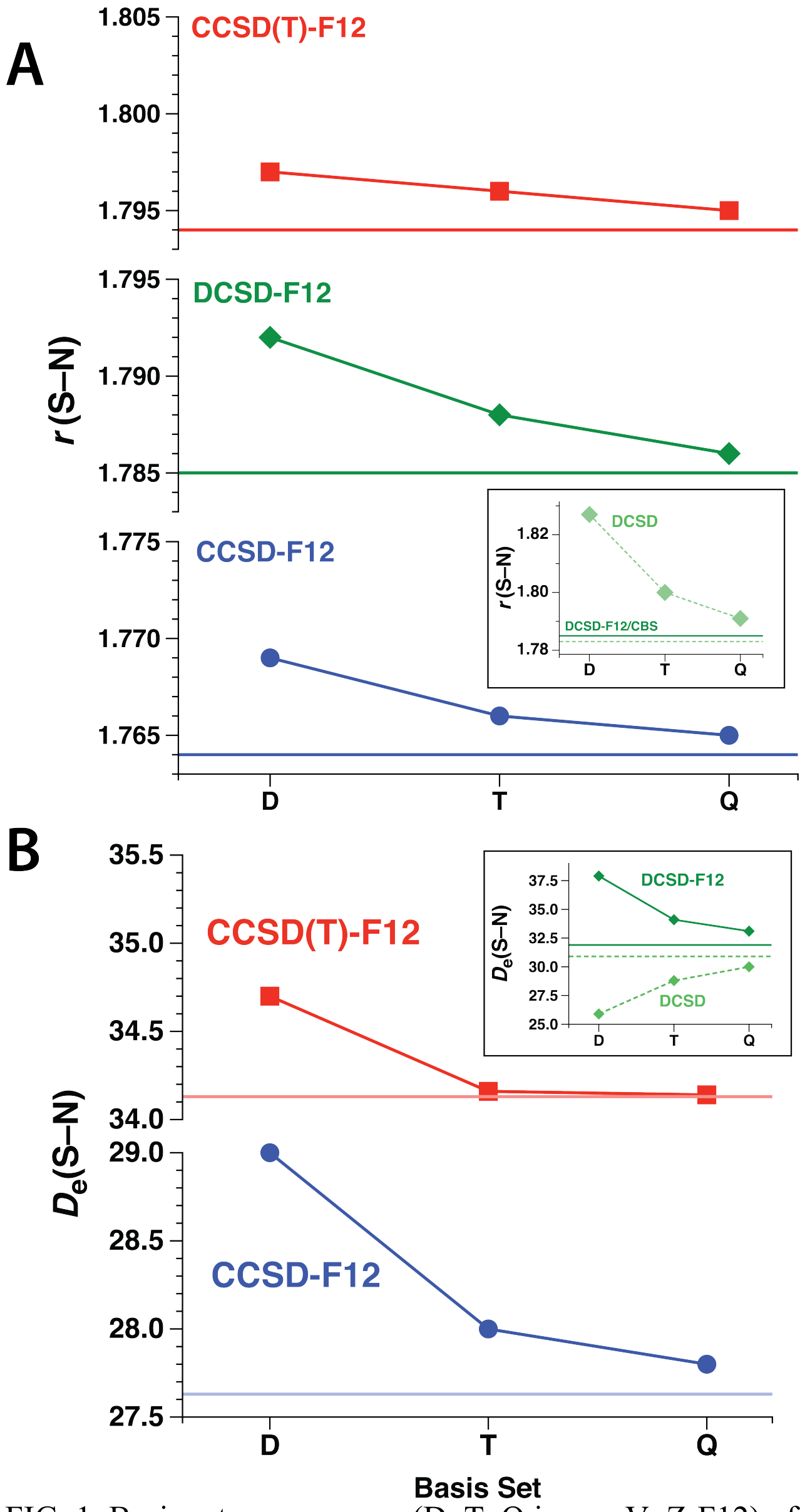

FIG. 1. Basis set convergence (D, T, Q in cc-pV*n*Z-F12) of the S–N bond lengths (A) and $D_e$(S–N) in cis-MeSNO (B). The horizontal lines show the respective $CBS_{(T\text{-}Q)}$ limits.

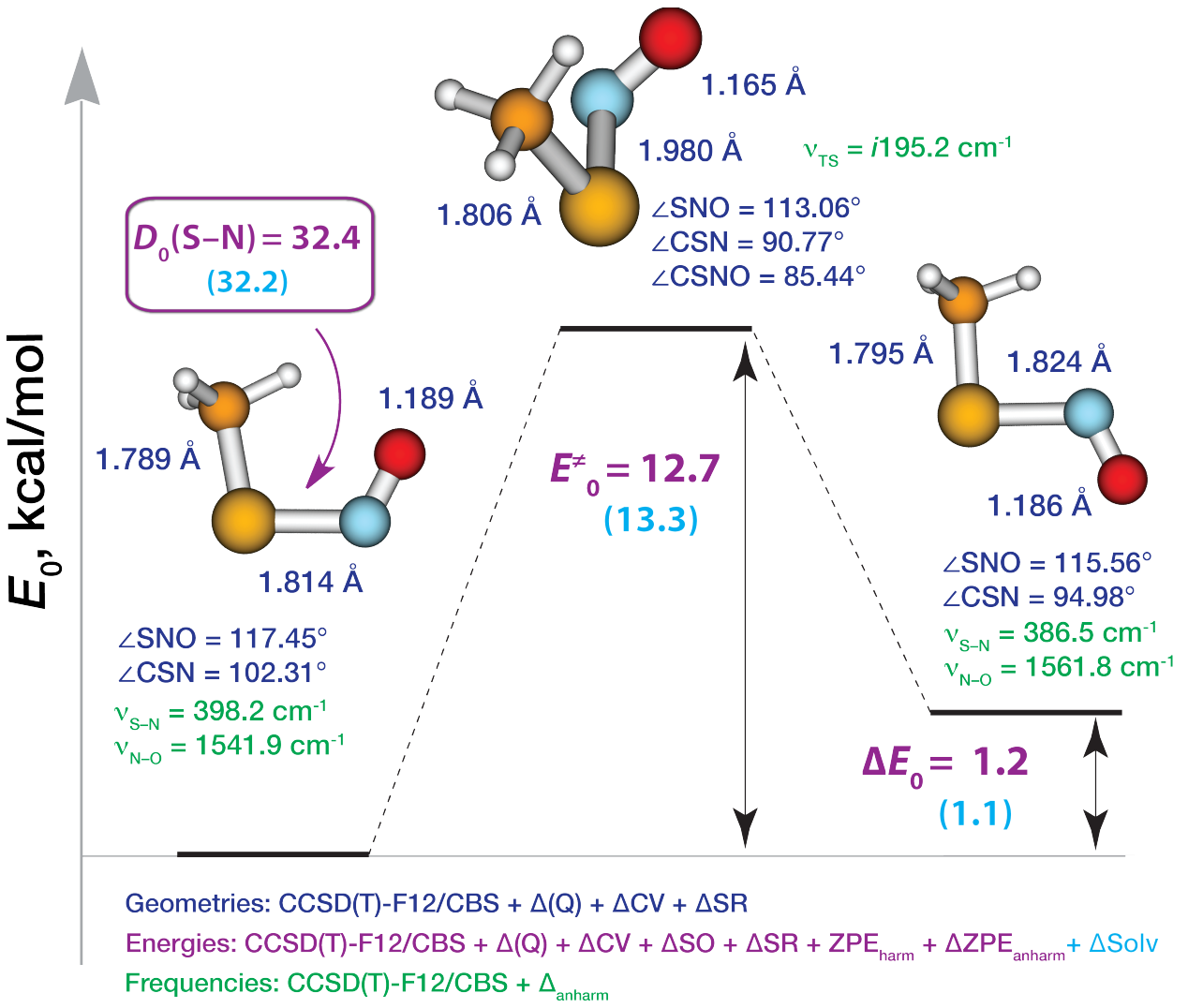

FIG. 2. Recommended *ab initio* geometric and energetic properties of cis-$CH_3SNO$ (left), $TS_{c\text{-}t}$ (middle), and trans-$CH_3SNO$ (right).

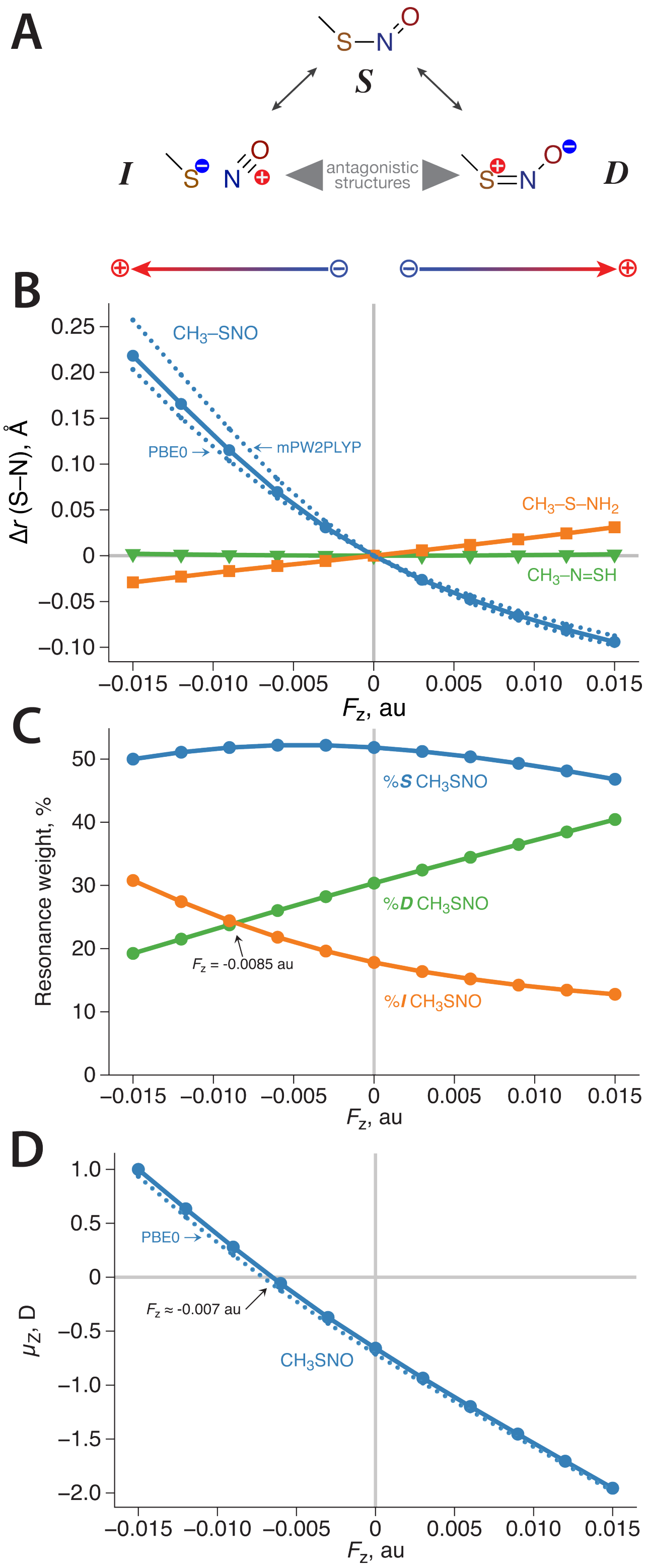

FIG. 3. Resonance description of the electronic structure of the -SNO group as a combination of conventional resonance structure ***S*** (single S–N bond), and antagonistic resonance structures ***D*** (double S–N bond) and ***I*** (ion pair) (A); EEF effects on the S–N bond lengths in $CH_3SNO$, $CH_3NS$ and $CH_3SNH_2$ molecules, calculated with CCSD(T)-F12/VDZ-F12, PBE0/def2-TZVPPD and mPW2PLYP/def2-TZVPPD methods (B); EEF effects on the resonance weights obtained from PBE0/def2-TZVPPD calculations of cis-$CH_3SNO$ optimized in EEF (C), dependence of the dipole moment projections $\mu_Z$ on the S–N vector calculated with CCSD(T)-F12/VDZ-F12 and PBE0/def2-TZVPPD for relaxed cis-$CH_3SNO$ geometries (D).

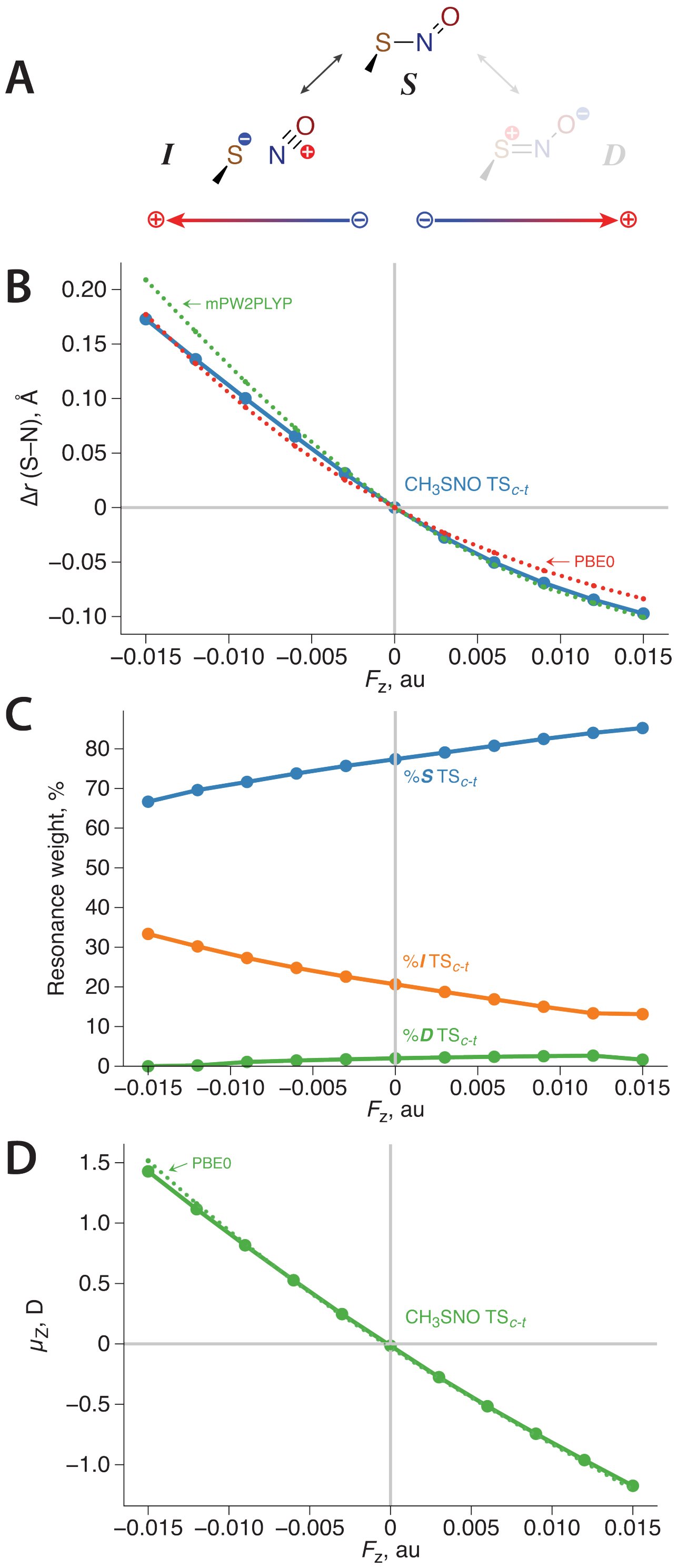

FIG. 4. Resonance description of the electronic structure of the -SNO group in $TS_{c\text{-}t}$ (A); EEF effects on the S–N bond lengths in $TS_{c\text{-}t}$, calculated with CCSD(T)-F12/VDZ-F12, PBE0/def2-TZVPPD and mPW2PLYP/def2-TZVPPD methods (B); EEF effects on the resonance weights obtained from PBE0/def2-TZVPPD calculations (C), dependence of the dipole moment projections $\mu_Z$ on the S–N vector calculated with CCSD(T)-F12/VDZ-F12 and PBE0/def2-TZVPPD for $TS_{c\text{-}t}$ geometries (D).

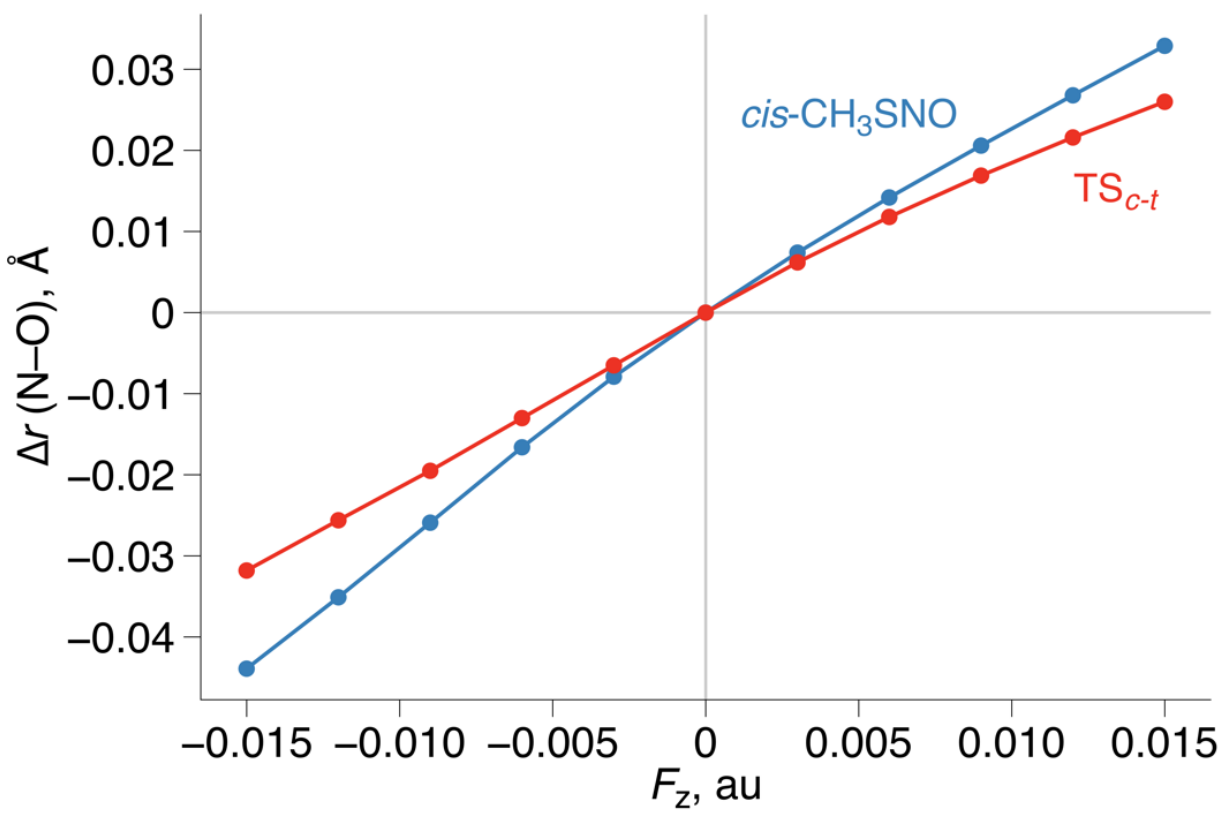

FIG. 5. EEF effect on the N–O bond lengths in cis-$CH_3SNO$ and $TS_{c\text{-}t}$, calculated with CCSD(T)-F12/VDZ-F12 methods.

# Toward Reliable Modeling of S-Nitrosothiol Chemistry: Structure and Properties of Methyl Thionitrite ($CH_3SNO$), an S-Nitrosocysteine Model

Dmitry G. Khomyakov and Qadir K. Timerghazin[a)]
*Chemistry Department, Marquette University, Milwaukee, Wisconsin, 53233, USA*

TABLE S1. Comparison of two CCSD(T)-F12/$CBS_{(T-Q)}$ extrapolation schemes for the energetic properties of $CH_3SNO$ (kcal/mol)

| $CBS_{(T-Q)}$ | $D_e$ | $\Delta E_e$ | $\Delta E_e^{\neq}$ |
|---|---|---|---|
| Equation (1) | 34.13 | 1.15 | 12.61 |
| Equation (2) | 34.20 | 1.15 | 12.61 |

TABLE S2. Cis-$CH_3SNO$, *ab initio* electronic energies and geometries

| Basis Set | *E*, Hartree | *r*(S–N), Å | *r*(N–O), Å | *r*(C–S), Å | ∠SNO, ° | ∠CSN, ° |
|---|---|---|---|---|---|---|
| CCSD(T)-F12 | | | | | | |
| VDZ-F12 | -567.369965 | 1.797 | 1.193 | 1.791 | 117.44 | 102.16 |
| VTZ-F12 | -567.401894 | 1.796 | 1.192 | 1.792 | 117.41 | 102.22 |
| VQZ-F12 | -567.407207 | 1.795 | 1.191 | 1.792 | 117.42 | 102.28 |
| $CBS_{(T-Q)}$ | -567.411084<br>*-567.410199** | 1.794 | 1.191 | 1.791 | 117.42 | 102.33 |
| CCSD-F12 | | | | | | |
| VDZ-F12 | -567.332994 | 1.769 | 1.186 | 1.790 | 117.86 | 103.05 |
| VTZ-F12 | -567.356558 | 1.766 | 1.185 | 1.790 | 117.90 | 103.18 |
| VQZ-F12 | -567.359266 | 1.765 | 1.184 | 1.790 | 117.91 | 103.25 |
| $CBS_{(T-Q)}$ | -567.361241 | 1.764 | 1.183 | 1.790 | 117.92 | 103.30 |
| DCSD-F12 | | | | | | |
| VDZ-F12 | -567.358628 | 1.792 | 1.192 | 1.792 | 117.51 | 102.52 |
| VTZ-F12 | -567.384701 | 1.788 | 1.19088 | 1.792 | 117.54 | 102.66 |
| VQZ-F12 | -567.388025 | 1.786 | 1.190 | 1.792 | 117.56 | 102.74 |
| $CBS_{(T-Q)}$ | -567.390451 | 1.785 | 1.190 | 1.791 | 117.569 | 102.79 |
| DCSD | | | | | | |
| VDZ-F12 | -567.205195 | 1.827 | 1.195 | 1.804 | 117.28 | 102.05 |
| VTZ-F12 | -567.323294 | 1.800 | 1.191 | 1.795 | 117.47 | 102.36 |
| VQZ-F12 | -567.358226 | 1.791 | 1.191 | 1.793 | 117.52 | 102.63 |
| $CBS_{(T-Q)}$ | -567.383717 | 1.783 | 1.191 | 1.792 | 117.560 | 102.82 |

* extrapolated with Schwenke-type CBS scheme (Equation 2)

TABLE S3. Trans-$CH_3SNO$, *ab initio* electronic energies and geometries

| Basis Set | *E*, Hartree | *r*(S–N), Å | *r*(N–O), Å | *r*(C–S), Å | ∠SNO, ° | ∠CSN, ° |
|---|---|---|---|---|---|---|
| CCSD(T)-F12 | | | | | | |
| VDZ-F12 | -567.368182 | 1.802 | 1.189 | 1.798 | 115.63 | 94.82 |
| VTZ-F12 | -567.400074 | 1.801 | 1.189 | 1.798 | 115.64 | 94.90 |
| VQZ-F12 | -567.405383 | 1.800 | 1.188 | 1.798 | 115.62 | 94.95 |
| $CBS_{(T-Q)}$ | -567.409258<br>*-567.408374** | 1.799 | 1.188 | 1.797 | 115.62 | 94.98 |
| CCSD-F12 | | | | | | |
| VDZ-F12 | -567.331792 | 1.776 | 1.184 | 1.795 | 115.11 | 95.66 |
| VTZ-F12 | -567.355356 | 1.774 | 1.183 | 1.795 | 115.13 | 95.79 |
| VQZ-F12 | -567.358062 | 1.773 | 1.182 | 1.795 | 115.13 | 95.85 |
| $CBS_{(T-Q)}$ | -567.360036 | 1.773 | 1.181 | 1.794 | 115.13 | 95.89 |
| DCSD-F12 | | | | | | |
| VDZ-F12 | -567.357324 | 1.797 | 1.189 | 1.798 | 115.31 | 95.07 |
| VTZ-F12 | -567.383395 | 1.794 | 1.189 | 1.798 | 115.31 | 95.20 |
| VQZ-F12 | -567.386718 | 1.793 | 1.188 | 1.797 | 115.31 | 95.26 |
| $CBS_{(T-Q)}$ | -567.389143 | 1.792 | 1.187 | 1.797 | 115.301 | 95.31 |
| DCSD | | | | | | |
| VDZ-F12 | -567.204246 | 1.828 | 1.193 | 1.810 | 115.43 | 94.49 |
| VTZ-F12 | -567.322081 | 1.805 | 1.189 | 1.801 | 115.41 | 94.97 |
| VQZ-F12 | -567.356938 | 1.796 | 1.189 | 1.799 | 115.30 | 95.19 |
| $CBS_{(T-Q)}$ | -567.382375 | 1.790 | 1.189 | 1.798 | 115.228 | 95.35 |

* extrapolated with Schwenke-type CBS scheme (Equation 2)

TABLE S4. Coupled cluster (CCSD(T)-F12) diagnostic values for $CH_3SNO$ isomers

| Basis set | Cis-$CH_3SNO$ | | Trans-$CH_3SNO$ | | Isomerization TS | |
|---|---|---|---|---|---|---|
| | T1 | D1 | T1 | D1 | T1 | D1 |
| VDZ-F12 | 0.025 | 0.081 | 0.025 | 0.079 | 0.020 | 0.073 |
| VTZ-F12 | 0.025 | 0.080 | 0.025 | 0.079 | 0.020 | 0.074 |
| VQZ-F12 | 0.025 | 0.080 | 0.025 | 0.079 | 0.020 | 0.076 |

TABLE S5. Cis-MeSNO S–N BDE ($D_e$), kcal/mol

| Basis Set/Method | DCSD | DCSD-F12 | CCSD-F12 | CCSD(T)-F12 |
|---|---|---|---|---|
| VDZ-F12 | 25.9 | 37.9 | 29.0 | 34.7 |
| VTZ-F12 | 28.8 | 34.8 | 28.0 | 34.2 |
| VQZ-F12 | 30.0 | 33.1 | 27.8 | 34.1 |
| $CBS_{(T-Q)}$ | 30.9 | 31.9 | 27.7 | 34.1 |

TABLE S6. Contribution of perturbative triple excitations to the properties of $CH_3SNO$ isomers

| Basis Set for CCSD-F12 and CCSD(T)-F12 | $D_e$(S–N), kcal/mol | $r$(S–N), Å | | Δ(T) in $CH_3SNO$ isomerization TS | |
|---|---|---|---|---|---|
| | | cis-$CH_3SNO$ | trans-$CH_3SNO$ | $r$(S–N), Å | $\Delta E_e^{\neq}$, kcal/mol |
| VDZ-F12 | 5.7 | 0.028 | 0.026 | 0.056 | 0.32 |
| VTZ-F12 | 6.2 | 0.030 | 0.027 | 0.060 | 0.84 |
| VQZ-F12 | 6.3 | 0.030 | 0.027 | 0.061 | 0.84 |

Contribution of perturbative quadruple excitations to the properties of $CH_3SNO$ isomers (calculated with the cc-pV(D+d)Z basis set)

| Method | $D_e$(S–N), kcal/mol | $r$(S–N), Å | | Isomerization TS | |
|---|---|---|---|---|---|
| | | cis-$CH_3SNO$ | trans-$CH_3SNO$ | $r$(S–N), Å | $\Delta E_e^{\neq}$, kcal/mol |
| CCSD(T) | 25.88 | 1.903 | 1.904 | 2.075 | 10.92 |
| CCSDT(Q) | 27.21 | 1.926 | 1.930 | 2.101 | 11.47 |
| Δ(Q) | 1.33 | 0.023 | 0.026 | 0.026 | 0.56 |

Contribution of perturbative quadruple excitations to the properties of $CH_3SNO$ isomers (calculated with the MIDI! basis set)

| Method | $D_e$(S–N), kcal/mol | $r$(S–N), Å | | Isomerization TS | |
|---|---|---|---|---|---|
| | | cis-$CH_3SNO$ | trans-$CH_3SNO$ | $r$(S–N), Å | $\Delta E_e^{\neq}$, kcal/mol |
| CCSD(T) | 31.49 | 1.922 | 1.920 | 1.903 | 11.50 |
| CCSDT(Q) | 32.43 | 1.938 | 1.940 | 1.926 | 11.96 |
| Δ(Q) | 0.95 | 0.016 | 0.020 | 0.023 | 0.46 |

TABLE S7. Solvent effects on $CH_3SNO$ energetic properties (PCM model, def2-TZVPPD basis set)

| Method | ΔSolv. ($H_2O$) | ΔSolv. (EtOEt) | ΔSolv. ($H_2O$) | ΔSolv. (EtOEt) | ΔSolv. ($H_2O$) | ΔSolv. (EtOEt) |
|---|---|---|---|---|---|---|
| | $D_0$(S–N), kcal/mol | | $\Delta E_0$, kcal/mol | | $\Delta E_0^{\neq}$, kcal/mol | |
| B3LYP | -0.22 | -0.19 | 0.01 | 0.00 | 0.51 | -0.17 |
| PBE0 | -0.22 | -0.18 | -0.02 | -0.02 | 0.51 | -0.14 |
| PBE0-GD3 | -0.21 | -0.17 | -0.01 | -0.01 | 0.55 | -0.14 |
| PBE0-1/3 | -0.13 | -0.12 | -0.05 | -0.04 | 0.61 | -0.10 |
| ωB97XD | -0.14 | -0.13 | -0.07 | -0.05 | 0.61 | -0.10 |
| B2PLYPD | -0.19 | -0.17 | 0.00 | 0.00 | 0.52 | -0.15 |
| mPW2PLYPD | -0.11 | -0.12 | -0.03 | -0.02 | 0.61 | -0.09 |
| Average | -0.18 | -0.15 | -0.02 | -0.02 | 0.57 | -0.13 |

TABLE S8. $CH_3SNO$ cis-trans relative energy $\Delta E_e$, kcal/mol

| Basis Set/Method | DCSD | DCSD-F12 | CCSD-F12 | CCSD(T)-F12 |
|---|---|---|---|---|
| VDZ-F12 | 0.6 | 0.8 | 0.8 | 1.1 |
| VTZ-F12 | 0.8 | 0.8 | 0.8 | 1.1 |
| VQZ-F12 | 0.8 | 0.8 | 0.8 | 1.1 |
| $CBS_{(T\text{-}Q)}$ | 0.8 | 0.8 | 0.8 | 1.1 |

TABLE S9. $CH_3SNO$ cis-trans isomerization TS, *ab initio* electronic energies and geometries

| Basis Set | *E*, Hartree | *r*(S–N), Å | *r*(N–O), Å | *r*(C–S), Å | ∠SNO, ° | ∠CSN, ° | ∠CSNO, ° |
|---|---|---|---|---|---|---|---|
| CCSD(T)-F12 | | | | | | | |
| VDZ-F12 | -567.349 854 | 1.956 | 1.169 | 1.808 | 113.08 | 90.71 | 85.63 |
| VTZ-F12 | -567.381835 | 1.957 | 1.168 | 1.809 | 113.06 | 90.69 | 85.49 |
| VQZ-F12 | -567.387124 | 1.956 | 1.168 | 1.809 | 113.07 | 90.79 | 85.43 |
| $CBS_{(T-Q)}$ | -567.390983<br>-567.390110* | 1.955 | 1.167 | 1.809 | 113.07 | 90.85 | 85.39 |
| CCSD-F12 | | | | | | | |
| VDZ-F12 | -567.314234 | 1.900 | 1.168 | 1.803 | 113.10 | 92.31 | 85.54 |
| VTZ-F12 | -567.337841 | 1.897 | 1.167 | 1.803 | 113.12 | 92.44 | 85.42 |
| VQZ-F12 | -567.340521 | 1.895 | 1.166 | 1.802 | 113.14 | 92.51 | 85.42 |
| $CBS_{(T-Q)}$ | -567.342477 | 1.894 | 1.166 | 1.802 | 113.15 | 92.57 | 85.42 |
| DCSD-F12 | | | | | | | |
| VDZ-F12 | -567.339521 | 1.936 | 1.172 | 1.807 | 113.01 | 91.46 | 85.57 |
| VTZ-F12 | -567.365616 | 1.933 | 1.171 | 1.807 | 113.00 | 91.52 | 85.43 |
| VQZ-F12 | -567.368908 | 1.931 | 1.170 | 1.807 | 113.01 | 91.62 | 85.44 |
| $CBS_{(T-Q)}$ | -567.371311 | 1.930 | 1.170 | 1.806 | 113.02 | 91.69 | 85.52 |
| DCSD | | | | | | | |
| VDZ-F12 | -567.187489 | 1.978 | 1.175 | 1.819 | 113.10 | 91.20 | 85.41 |
| VTZ-F12 | -567.304740 | 1.946 | 1.172 | 1.810 | 112.98 | 91.33 | 85.43 |
| VQZ-F12 | -567.339289 | 1.935 | 1.171 | 1.809 | 113.01 | 91.52 | 85.43 |
| $CBS_{(T-Q)}$ | -567.364501 | 1.927 | 1.171 | 1.808 | 113.03 | 91.66 | 85.42 |

* extrapolated with Schwenke-type CBS scheme (Equation 2)

TABLE S10. $CH_3SNO$ cis-trans isomerization barrier $\Delta E_e^{\neq}$, kcal/mol

| Basis Set/Method | DCSD | DCSD-F12 | CCSD-F12 | CCSD(T)-F12 |
|---|---|---|---|---|
| VDZ-F12 | 11.1 | 11.99 | 11.8 | 12.1 |
| VTZ-F12 | 11.6 | 11.98 | 11.7 | 12.6 |
| VQZ-F12 | 11.9 | 12.00 | 11.8 | 12.6 |
| $CBS_{(T-Q)}$ | 12.1 | 12.01 | 11.8 | 12.6 |

TABLE S11. Cis-$CH_3SNO$ *ab initio* harmonic vibrational frequencies

| Mode | CCSD(T)-F12/VDZ-F12 | CCSD(T)-F12/VTZ-F12 | CCSD(T)-F12/VQZ-F12 | CCSD(T)-F12/CBS (T-Q) |
|---|---|---|---|---|
| 1 A" | 92.0 | 88.4 | 86.2 | 84.6 |
| 2 A' | 281.2 | 281.0 | 280.5 | 280.1 |
| 3 A" | 290.2 | 290.8 | 290.9 | 291.0 |
| 4 A' (S–N) | 400.6 | 399.1 | 399.5 | 399.7 |
| 5 A' | 663.3 | 662.4 | 662.4 | 662.4 |
| 6 A' | 756.4 | 754.0 | 754.2 | 754.3 |
| 7 A' | 964.0 | 963.2 | 962.8 | 962.5 |
| 8 A" | 973.4 | 970.9 | 970.8 | 970.8 |
| 9 A' | 1340.2 | 1335.6 | 1334.8 | 1334.3 |
| 10 A" | 1480.4 | 1475.4 | 1476.4 | 1477.2 |
| 11 A' | 1485.3 | 1480.3 | 1480.6 | 1480.8 |
| 12 A' (N–O) | 1571.5 | 1572.2 | 1573.9 | 1575.1 |
| 13 A' | 3032.1 | 3027.8 | 3028.2 | 3028.4 |
| 14 A' | 3128.9 | 3126.1 | 3126.8 | 3127.3 |
| 15 A" | 3163.6 | 3160.8 | 3158.7 | 3157.2 |
| $ZPE_{harm}$, kcal/mol | 28.1 | 28.0 | 28.0 | 28.0 |

TABLE S12. Trans-$CH_3SNO$ *ab initio* harmonic vibrational frequencies

| Mode | CCSD(T)-F12/VDZ-F12 | CCSD(T)-F12/VTZ-F12 | CCSD(T)-F12/VQZ-F12 | CCSD(T)-F12/CBS (T-Q) |
|---|---|---|---|---|
| 1 A" | 121.7 | 119.7 | 119.6 | 119.5 |
| 2 A' | 227.9 | 227.7 | 227.6 | 227.5 |
| 3 A" | 236.1 | 235.7 | 235.6 | 235.6 |
| 4 A' (S–N) | 394.2 | 393.5 | 393.9 | 394.2 |
| 5 A' | 670.3 | 669.3 | 669.5 | 669.6 |
| 6 A' | 753.2 | 751.0 | 751.5 | 751.8 |
| 7 A' | 972.6 | 971.0 | 994.1 | 1011.0 |
| 8 A" | 996.5 | 994.4 | 970.8 | 953.5 |
| 9 A' | 1358.5 | 1353.5 | 1353.0 | 1352.6 |
| 10 A" | 1471.9 | 1468.5 | 1468.4 | 1468.3 |
| 11 A' | 1498.3 | 1493.1 | 1493.9 | 1494.5 |
| 12 A' (N–O) | 1590.7 | 1590.9 | 1592.4 | 1593.5 |
| 13 A' | 3049.3 | 3044.7 | 3045.6 | 3046.2 |
| 14 A' | 3142.9 | 3140.1 | 3141.4 | 3142.3 |
| 15 A" | 3162.2 | 3159.9 | 3154.5 | 3150.6 |
| $ZPE_{harm}$, kcal/mol | 28.1 | 28.0 | 28.0 | 28.0 |

TABLE S13. DFT anharmonic correction to vibrational frequencies of cis-$CH_3SNO$ and trans-$CH_3SNO$ (calculated with the def2-TZVPPD basis set)

| Mode | mPW2-PLYPD | PBE0 | PBE0-GD3 | mPW2-PLYPD | PBE0 | PBE0-GD3 |
|---|---|---|---|---|---|---|
| | Cis-$CH_3SNO$ | | | Trans-$CH_3SNO$ | | |
| 1 A" | -39.6 | -47.6 | -29.3 | -17.3 | -19.4 | -18.8 |
| 2 A' | -10.0 | -11.3 | -5.2 | -6.9 | -7.6 | -7.1 |
| 3 A" | -4.8 | -2.8 | -5.4 | -11.1 | -15.6 | -17.7 |
| 4 A' (S–N) | -1.5 | -4.1 | -2.8 | -7.6 | -7.6 | -6.7 |
| 5 A' | -9.5 | -9.3 | -9.2 | -9.5 | -8.5 | -8.8 |
| 6 A' | -16.4 | -17.7 | -17.1 | -14.7 | -14.3 | -14.3 |
| 7 A' | -22.8 | -18.5 | -15.3 | -16.3 | -14.9 | -13.8 |
| 8 A" | -12.8 | -19.6 | -18.4 | -17.8 | -18.7 | -18.9 |
| 9 A' | -35.2 | -28.7 | -28.7 | -32.9 | -29.5 | -29.4 |
| 10 A" | -41.3 | -38.6 | -39.5 | -41.7 | -41.8 | -42.8 |
| 11 A' | -44.7 | -40.0 | -41.2 | -42.4 | -42.0 | -42.6 |
| 12 A' (N–O) | -33.2 | -25.7 | -27.0 | -31.8 | -27.7 | -28.1 |
| 13 A' | -96.8 | -105.3 | -105.4 | -98.2 | -106.7 | -107.0 |
| 14 A' | -137.3 | -137.6 | -135.3 | -138.9 | -136.5 | -136.6 |
| 15 A" | -145.3 | -145.4 | -143.5 | -144.2 | -141.2 | -141.1 |
| $\Delta ZPE_{anharm}$, kcal/mol | -0.39 | -0.38 | -0.37 | -0.38 | -0.37 | -0.37 |

TABLE S14. Performance of the DFT methods (with def2-TZVPPD) vs. truncated FPD scheme for trans-$CH_3SNO$

| Method | $r$(S–N), Å | $r$(N–O), Å | $r$(C–S), Å | ∠SNO, ° | $\Delta E_0$, kcal/mol |
|---|---|---|---|---|---|
| CCSD(T)-F12/ $CBS_{(T-Q)}$+ Δ(Q)+ΔCV+$ZPE_{harm}$ | 1.820 | 1.186 | 1.793 | 115.62 | 1.13 |
| B3LYP | 1.827 | 1.177 | 1.805 | 116.50 | 0.82 |
| PBE0 | 1.790 | 1.175 | 1.788 | 116.45 | 1.10 |
| PBE0-GD3 | 1.791 | 1.175 | 1.789 | 116.45 | 1.34 |
| PBE0-1/3 | 1.772 | 1.172 | 1.784 | 116.23 | 1.06 |
| ωB97XD | 1.779 | 1.177 | 1.793 | 115.84 | 1.05 |
| B2PLYP | 1.819 | 1.186 | 1.800 | 116.50 | 1.10 |
| B2PLYPD | 1.821 | 1.186 | 1.802 | 116.46 | 1.35 |
| mPW2PLYP | 1.803 | 1.183 | 1.798 | 116.31 | 1.11 |
| mPW2PLYPD | 1.804 | 1.183 | 1.799 | 116.29 | 1.29 |

TABLE S15. Performance of the DFT methods (with def2-SV(P)+d basis set) vs. truncated FPD scheme for cis-$CH_3SNO$

| Method | $r$(S–N), Å | $r$(N–O), Å | $r$(C–S), Å | ∠SNO, ° | $D_0$(S–N), kcal/mol |
|---|---|---|---|---|---|
| CCSD(T)-F12/ $CBS_{(T-Q)}$+ Δ(Q)+ΔCV+$ZPE_{harm}$ | 1.810 | 1.189 | 1.787 | 117.51 | 32.67 |
| B3LYP | 1.830 | 1.181 | 1.797 | 117.68 | 29.64 |
| PBE0 | 1.795 | 1.177 | 1.780 | 117.85 | 31.58 |
| PBE0-GD3 | 1.794 | 1.177 | 1.782 | 117.92 | 32.27 |
| PBE0-1/3 | 1.769 | 1.174 | 1.778 | 118.17 | 28.28 |
| ωB97XD | 1.774 | 1.181 | 1.788 | 118.06 | 28.92 |
| B2PLYPD | 1.826 | 1.187 | 1.794 | 117.72 | 27.57 |
| mPW2PLYPD | 1.807 | 1.185 | 1.792 | 117.89 | 26.67 |

TABLE S16. Performance of the DFT methods (with def2-SV(P)+d basis set) vs. truncated FPD scheme for trans-$CH_3SNO$

| Method | $r$(S–N), Å | $r$(N–O), Å | $r$(C–S), Å | ∠SNO, ° | $\Delta E_0$, kcal/mol |
|---|---|---|---|---|---|
| CCSD(T)-F12/ $CBS_{(T-Q)}$+ Δ(Q)+ΔCV+$ZPE_{harm}$ | 1.820 | 1.186 | 1.793 | 115.62 | 1.13 |
| B3LYP | 1.847 | 1.175 | 1.802 | 116.83 | 1.35 |
| PBE0 | 1.813 | 1.171 | 1.786 | 116.85 | 1.59 |
| PBE0-GD3 | 1.814 | 1.171 | 1.787 | 116.84 | 1.82 |
| PBE0-1/3 | 1.790 | 1.169 | 1.783 | 116.59 | 1.55 |
| ωB97XD | 1.791 | 1.176 | 1.792 | 116.11 | 1.54 |
| B2PLYPD | 1.843 | 1.181 | 1.800 | 116.91 | 1.74 |
| mPW2PLYPD | 1.824 | 1.179 | 1.797 | 116.70 | 1.73 |

TABLE S17. Performance of the DFT methods (with def2-SV(P)+d basis set) vs. truncated FPD scheme for $CH_3SNO$ cis-trans isomerization TS

| Method | $r$(S–N), Å | $r$(N–O), Å | $r$(C–S), Å | ∠SNO, ° | ∠CSNO, ° | $\Delta E_0^{\neq}$, kcal/mol |
|---|---|---|---|---|---|---|
| CCSD(T)-F12/ $CBS_{(T-Q)}$+ Δ(Q)+ΔCV+$ZPE_{harm}$ | 1.949 | 1.166 | 1.805 | 113.09 | 85.42 | 12.71 |
| B3LYP | 2.009 | 1.157 | 1.815 | 113.57 | 86.14 | 14.42 |
| PBE0 | 1.966 | 1.153 | 1.799 | 113.49 | 85.77 | 15.16 |
| PBE0-GD3 | 1.966 | 1.153 | 1.799 | 113.51 | 86.01 | 15.19 |
| PBE0-1/3 | 1.936 | 1.151 | 1.794 | 113.48 | 85.74 | 14.72 |
| ωB97XD | 1.942 | 1.158 | 1.803 | 113.30 | 86.55 | 14.02 |
| B2PLYPD | 2.010 | 1.162 | 1.812 | 113.49 | 86.22 | 14.04 |
| mPW2PLYPD | 1.985 | 1.161 | 1.808 | 113.44 | 86.24 | 13.99 |

TABLE S18. Cis-$CH_3SNO$ DFT harmonic vibrational frequencies (with def2-TZVPPD basis set)

| Mode | B3LYP | PBE0 | PBE0-1/3 | PBE0-GD3 | ωB97XD | B2-PLYPD | mPW2-PLYPD |
|---|---|---|---|---|---|---|---|
| 1 A" | 105.4 | 94.0 | 97.7 | 105.7 | 77.7 | 103.7 | 101.6 |
| 2 A' | 252.5 | 265.7 | 272.3 | 264.6 | 256.4 | 261.3 | 266.3 |
| 3 A" | 298.4 | 311.8 | 314.4 | 311.1 | 303.4 | 301.3 | 303.1 |
| 4 A' (S–N) | 376.6 | 426.6 | 452.7 | 424.1 | 444.4 | 370.1 | 388.4 |
| 5 A' | 658.0 | 677.7 | 684.4 | 677.1 | 673.8 | 662.8 | 666.8 |
| 6 A' | 728.9 | 763.2 | 773.9 | 759.9 | 757.3 | 738.6 | 744.8 |
| 7 A' | 959.3 | 952.4 | 964.5 | 953.2 | 960.2 | 963.0 | 969.9 |
| 8 A" | 962.5 | 958.2 | 971.9 | 958.2 | 971.6 | 967.4 | 974.9 |
| 9 A' | 1333.5 | 1320.7 | 1337.3 | 1321.9 | 1337.6 | 1338.0 | 1346.5 |
| 10 A" | 1467.8 | 1457.1 | 1473.1 | 1456.5 | 1472.4 | 1475.2 | 1484.0 |
| 11 A' | 1472.2 | 1461.8 | 1478.2 | 1461.6 | 1476.7 | 1479.1 | 1488.2 |
| 12 A' (N–O) | 1627.0 | 1666.0 | 1693.6 | 1666.5 | 1662.8 | 1553.7 | 1578.3 |
| 13 A' | 3023.3 | 3037.1 | 3065.3 | 3031.7 | 3046.1 | 3023.2 | 3043.8 |
| 14 A' | 3109.7 | 3135.9 | 3162.1 | 3129.6 | 3142.0 | 3123.9 | 3141.4 |
| 15 A" | 3140.3 | 3171.3 | 3194.5 | 3164.7 | 3175.4 | 3161.2 | 3176.4 |
| $ZPE_{harm}$, kcal/mol | 27.9 | 28.2 | 28.5 | 28.1 | 28.2 | 27.9 | 28.1 |

TABLE S19. Cis-$CH_3SNO$ DFT harmonic vibrational frequencies (with def2-SV(P)+d basis set)

| Mode | B3LYP | PBE0 | PBE0-1/3 | PBE0-GD3 | ωB97XD | B2-PLYPD | mPW2-PLYPD |
|---|---|---|---|---|---|---|---|
| 1 A" | 147.2 | 149.7 | 137.1 | 145.7 | 117.8 | 142.5 | 136.4 |
| 2 A' | 268.7 | 281.3 | 288.1 | 279.9 | 284.5 | 276.9 | 282.6 |
| 3 A" | 303.5 | 316.1 | 319.5 | 316.0 | 310.2 | 304.7 | 307.2 |
| 4 A' (S–N) | 369.3 | 409.7 | 441.9 | 408.9 | 440.1 | 359.2 | 380.9 |
| 5 A' | 671.8 | 690.3 | 696.7 | 689.9 | 688.0 | 677.9 | 682.2 |
| 6 A' | 740.9 | 771.5 | 780.3 | 768.6 | 769.1 | 751.4 | 757.8 |
| 7 A' | 966.2 | 969.0 | 981.3 | 967.5 | 977.7 | 978.3 | 984.5 |
| 8 A" | 969.5 | 971.3 | 982.4 | 970.1 | 977.8 | 982.4 | 988.3 |
| 9 A' | 1332.3 | 1332.0 | 1347.6 | 1331.9 | 1346.3 | 1353.2 | 1360.0 |
| 10 A" | 1456.7 | 1457.1 | 1473.7 | 1456.3 | 1467.3 | 1473.8 | 1482.1 |
| 11 A' | 1457.7 | 1457.5 | 1473.8 | 1456.4 | 1468.5 | 1475.0 | 1483.2 |
| 12 A' (N–O) | 1714.5 | 1757.3 | 1780.7 | 1756.3 | 1745.0 | 1653.8 | 1674.9 |
| 13 A' | 2996.6 | 3013.8 | 3050.0 | 3011.3 | 3024.0 | 3005.2 | 3028.5 |
| 14 A' | 3097.6 | 3128.6 | 3163.3 | 3125.5 | 3136.4 | 3118.2 | 3138.9 |
| 15 A" | 3132.7 | 3169.5 | 3200.8 | 3166.4 | 3172.5 | 3158.6 | 3176.9 |
| $ZPE_{harm}$, kcal/mol | 28.1 | 28.4 | 28.8 | 28.4 | 28.5 | 28.2 | 28.4 |

TABLE S20. Trans-$CH_3SNO$ DFT harmonic vibrational frequencies (with def2-TZVPPD basis set)

| Mode | B3LYP | PBE0 | PBE0-1/3 | PBE0-GD3 | ωB97XD | B2-PLYPD | mPW2-PLYPD |
|---|---|---|---|---|---|---|---|
| 1 A" | 130.0 | 122.0 | 125.6 | 131.4 | 117.3 | 131.3 | 129.5 |
| 2 A' | 229.6 | 238.9 | 243.0 | 237.8 | 235.9 | 229.6 | 235.0 |
| 3 A" | 236.1 | 243.1 | 245.4 | 242.4 | 241.3 | 239.3 | 239.0 |
| 4 A' (S–N) | 372.5 | 412.9 | 432.2 | 409.5 | 424.0 | 368.0 | 384.5 |
| 5 A' | 662.2 | 685.2 | 695.9 | 683.6 | 687.2 | 664.9 | 671.5 |
| 6 A' | 726.8 | 762.0 | 773.7 | 758.8 | 757.0 | 735.3 | 742.2 |
| 7 A' | 965.5 | 960.0 | 972.7 | 959.5 | 970.6 | 974.3 | 980.3 |
| 8 A" | 986.7 | 984.4 | 997.8 | 984.3 | 996.2 | 992.9 | 1000.2 |
| 9 A' | 1349.3 | 1339.2 | 1356.4 | 1339.0 | 1358.7 | 1358.5 | 1366.3 |
| 10 A" | 1460.4 | 1449.5 | 1465.3 | 1449.4 | 1465.3 | 1468.0 | 1476.5 |
| 11 A' | 1485.0 | 1476.0 | 1492.2 | 1475.3 | 1490.5 | 1493.5 | 1502.2 |
| 12 A' (N–O) | 1657.0 | 1696.4 | 1722.2 | 1698.3 | 1688.5 | 1582.3 | 1604.3 |
| 13 A' | 3039.2 | 3057.3 | 3081.9 | 3051.0 | 3061.4 | 3044.2 | 3063.0 |
| 14 A' | 3123.3 | 3151.6 | 3177.0 | 3145.2 | 3154.8 | 3139.5 | 3157.1 |
| 15 A" | 3140.4 | 3170.6 | 3193.7 | 3164.5 | 3172.4 | 3160.8 | 3176.9 |
| $ZPE_{harm}$, kcal/mol | 28.0 | 28.2 | 28.6 | 28.2 | 28.3 | 28.0 | 28.2 |

TABLE S21. Trans-$CH_3SNO$ DFT harmonic vibrational frequencies (with def2-SV(P)+d basis set)

| Mode | B3LYP | PBE0 | PBE0-1/3 | PBE0-GD3 | ωB97XD | B2-PLYPD | mPW2-PLYPD |
|---|---|---|---|---|---|---|---|
| 1 A" | 146.9 | 150.3 | 142.4 | 148.6 | 135.1 | 146.5 | 143.7 |
| 2 A' | 235.9 | 245.1 | 251.5 | 244.0 | 245.4 | 235.0 | 240.6 |
| 3 A" | 247.8 | 255.3 | 253.4 | 254.5 | 248.6 | 251.3 | 250.3 |
| 4 A' (S–N) | 369.8 | 402.6 | 424.4 | 400.7 | 422.8 | 362.4 | 379.9 |
| 5 A' | 669.8 | 688.8 | 698.8 | 688.0 | 692.8 | 673.3 | 680.0 |
| 6 A' | 739.4 | 771.0 | 781.7 | 768.3 | 769.3 | 750.0 | 757.8 |
| 7 A' | 971.0 | 972.4 | 984.4 | 970.5 | 979.9 | 986.8 | 992.5 |
| 8 A" | 988.0 | 993.0 | 1005.3 | 991.4 | 1001.1 | 1001.2 | 1007.6 |
| 9 A' | 1344.1 | 1344.5 | 1361.4 | 1343.7 | 1359.8 | 1367.2 | 1374.0 |
| 10 A" | 1446.8 | 1446.6 | 1462.7 | 1445.8 | 1457.6 | 1463.7 | 1471.7 |
| 11 A' | 1468.7 | 1470.8 | 1487.7 | 1469.8 | 1481.4 | 1487.2 | 1495.5 |
| 12 A' (N–O) | 1755.8 | 1797.9 | 1818.7 | 1797.6 | 1779.3 | 1695.8 | 1714.4 |
| 13 A' | 3015.9 | 3037.5 | 3069.5 | 3034.3 | 3041.1 | 3027.0 | 3047.9 |
| 14 A' | 3114.1 | 3146.5 | 3180.8 | 3143.6 | 3152.4 | 3135.7 | 3155.7 |
| 15 A" | 3131.6 | 3167.9 | 3199.7 | 3165.0 | 3170.2 | 3157.0 | 3175.4 |
| $ZPE_{harm}$, kcal/mol | 28.1 | 28.4 | 28.8 | 28.4 | 28.5 | 28.2 | 28.4 |

TABLE S22. The S–N bond force constants, obtained via fitting of the S–N bond energy profiles in $CH_3SNO$ to the harmonic potential, $E=k\Delta x^2$.

| Method/Basis set | $k$, mDyne/Å | $R^2$ | $r(S–N)_{min}$, Å |
|---|---|---|---|
| | | cis-$CH_3SNO$ | |
| CCSD/cc-pV(D+d)Z | 0.78313 | 0.9683 | 1.868 |
| DCSD/cc-pV(D+d)Z | 0.67259 | 0.9976 | 1.900 |
| CCSD(T)/cc-pV(D+d)Z | 0.65873 | 0.8909 | 1.903 |
| CCSDT/cc-pV(D+d)Z | 0.62930 | 0.9787 | 1.912 |
| CCSDT(Q)/cc-pV(D+d)Z | 0.55738 | 0.9603 | 1.926 |
| | | $TS_{c\text{-}t}$ | |
| CCSD/cc-pV(D+d)Z | 0.53636 | 0.9931 | 1.998 |
| DCSD/cc-pV(D+d)Z | 0.43438 | 0.9813 | 2.052 |
| CCSD(T)/cc-pV(D+d)Z | 0.37117 | 0.9279 | 2.075 |
| CCSDT/cc-pV(D+d)Z | 0.36307 | 0.9474 | 2.082 |
| CCSDT(Q)/cc-pV(D+d)Z | 0.33084 | 0.9771 | 2.101 |

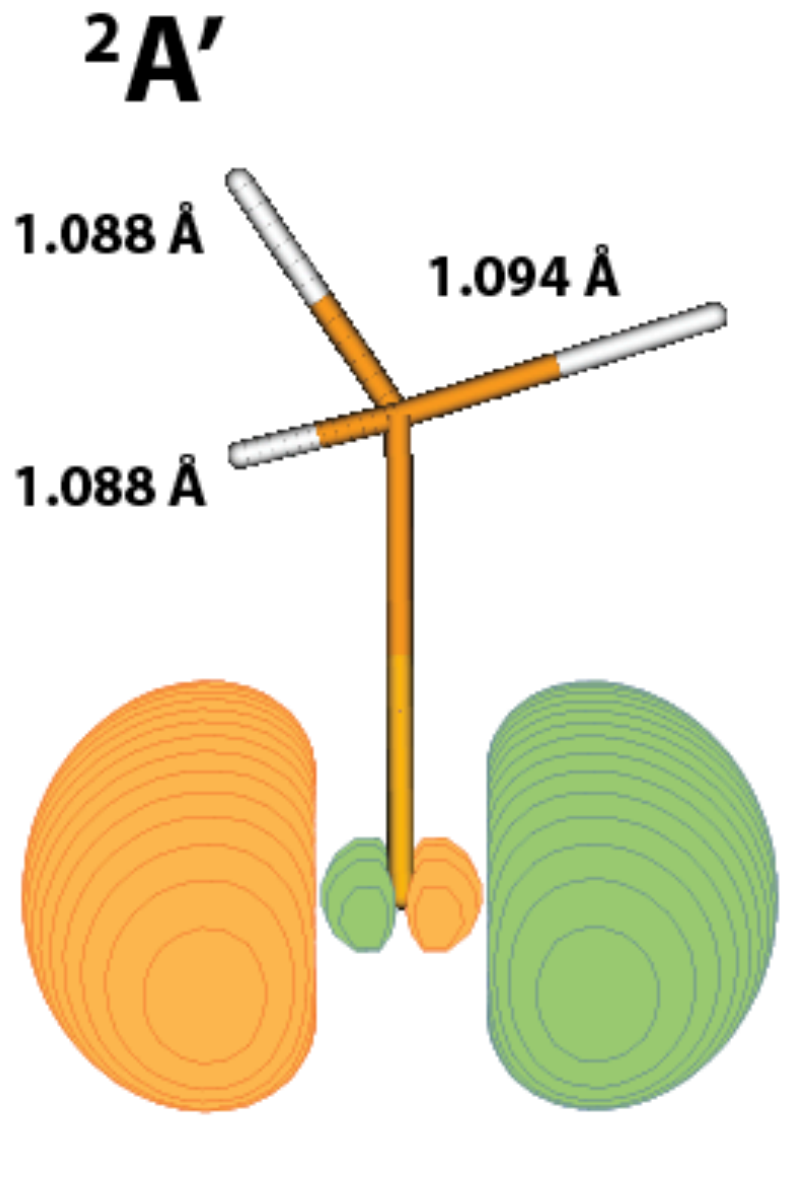

FIG. S1. Non-degenerate $^2A'$ and $^2A''$ electronic states of MeS• radical after Jahn-Teller distortion (Cs symmetry). C–H bond distances are calculated at the CCSD(T)-F12/VQZ-F12 level of theory.

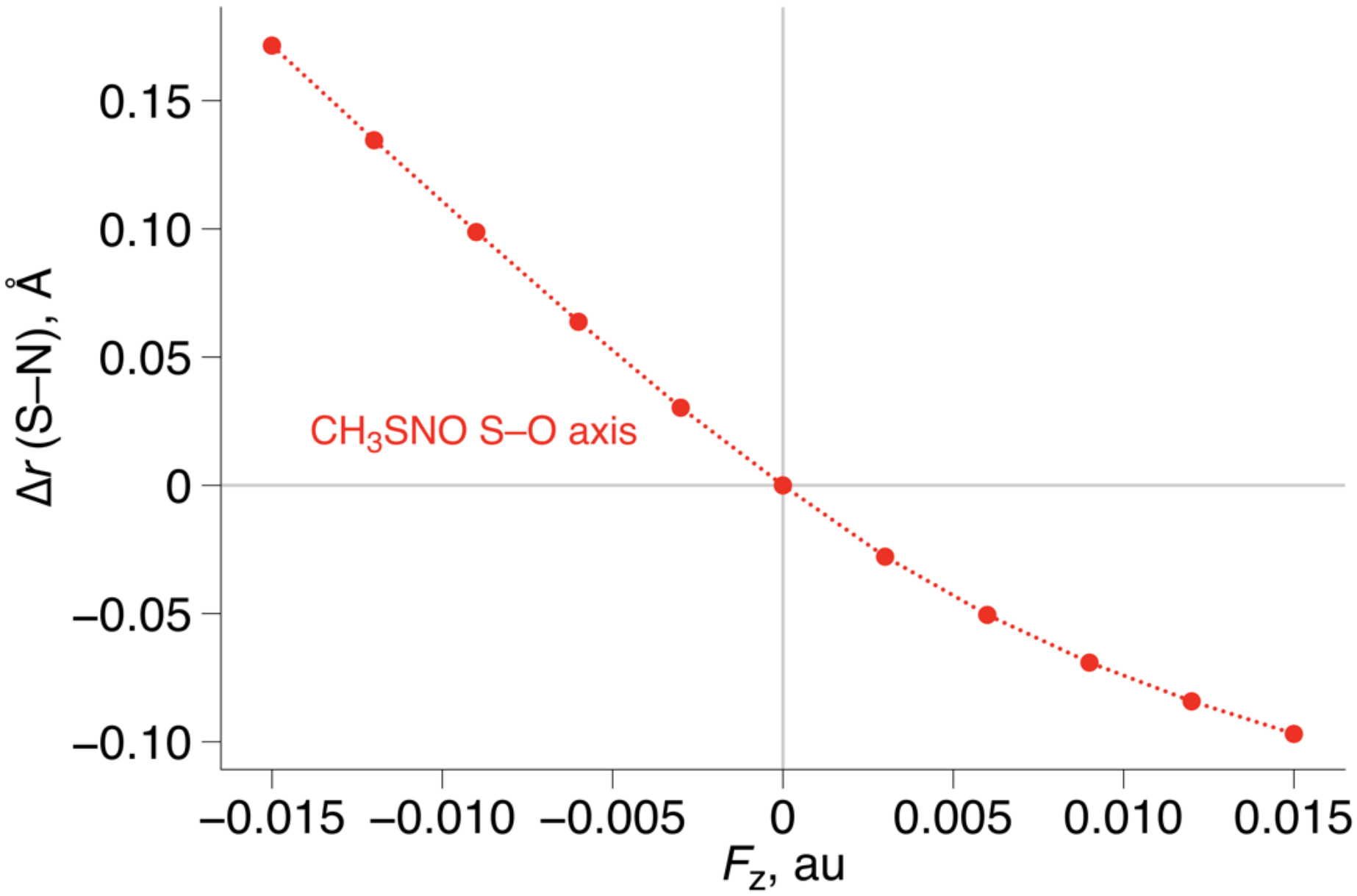

FIG. S2. EEF effect on the S–N bond length in cis-$CH_3SNO$, EEFs oriented along the S-O vector.

Toward Reliable Modeling of S–Nitrosothiol Chemistry: Structure and Properties of Methyl Thionitrite (CH3SNO), an S–Nitrosocysteine Model

Dmitry G. Khomyakov and Qadir K. Timerghazin
Chemistry Department, Marquette University, Milwaukee, Wisconsin, 53233, USA

Cartesian Coordinates and Energies

CCSD(T)–F12a/VQZ–F12a Geometries and Energies

Cis–CH3SNO:
CCSD(T)–F12A/VQZ–F12; E = –567.407207 Hartree

| | | | |
|---|---|---|---|
| C | 1.5516649662 | 0.0404651186 | 0.0000000000 |
| S | 0.0008581743 | 0.9375271884 | 0.0000000000 |
| H | 2.1216706328 | 0.2818320777 | 0.8935492740 |
| H | 1.2961372625 | –1.0234212292 | 0.0000000000 |
| H | 2.1216706328 | 0.2818320777 | –0.8935492740 |
| N | –1.2077020234 | –0.3893614113 | 0.0000000000 |
| O | –0.7952936452 | –1.5069838220 | 0.0000000000 |

Trans–CH3SNO:
CCSD(T)–F12A/VQZ–F12 E = –567.405383 Hartree

| | | | |
|---|---|---|---|
| C | 1.7613517307 | 0.4169589956 | 0.0000000000 |
| S | –0.0020377956 | 0.7670548341 | 0.0000000000 |
| H | 2.2257572477 | 0.8299448185 | 0.8915698914 |
| H | 1.8615611956 | –0.6698260778 | 0.0000000000 |
| H | 2.2257572477 | 0.8299448185 | –0.8915698914 |
| N | –0.5035429766 | –0.9617040027 | 0.0000000000 |
| O | –1.6756526497 | –1.1567353864 | 0.0000000000 |

Isomerization TS:
CCSD(T)–F12A/VQZ–F12 E = –567.387123 Hartree

| | | | |
|---|---|---|---|
| C | –1.0505530096 | 0.0284314491 | –1.3184526979 |
| S | 0.6467359054 | –0.0841764195 | –0.7031503056 |
| N | 0.0763118021 | 0.5339576380 | 1.0622964886 |
| O | –0.3131481123 | –0.3293501418 | 1.7455416848 |
| H | –1.6877455072 | –0.7130586908 | –0.8397125340 |
| H | –1.4468216085 | 1.0318146755 | –1.1657896541 |
| H | –1.0074011970 | –0.1722928156 | –2.3875705213 |

Vibrational frequencies, cm–1
3041.43
3144.85
3131.30
1676.02
1487.13
1472.82
1347.75
976.30
953.34
726.92
595.21
307.65
200.98
146.93
–194.48

ZPE, [1/CM]:  9604.31
ZPE, [H]:  0.043760

Normal modes:

| | | | | | |
|---|---|---|---|---|---|
| freq. | 3041.43 | 3144.85 | 3131.30 | 1676.02 | 1487.13 |
| C 0 | −0.04192 | 0.00713 | −0.01008 | −0.00011 | −0.00011 |
| C 1 | 0.02999 | −0.00139 | −0.08337 | −0.00013 | −0.00005 |
| C 2 | −0.00798 | −0.08975 | 0.00009 | 0.00004 | 0.00016 |
| S 0 | −0.00205 | −0.00114 | −0.00199 | −0.00402 | 0.00309 |
| S 1 | −0.01887 | 0.01391 | 0.03155 | −0.00098 | 0.00084 |
| S 2 | 0.00864 | 0.03397 | −0.01013 | 0.00044 | 0.00312 |
| N 0 | −0.08829 | 0.00480 | −0.04341 | −0.00867 | −0.00068 |
| N 1 | 0.02211 | 0.09426 | −0.02785 | −0.01055 | −0.03434 |
| N 2 | 0.05503 | −0.03620 | −0.08869 | −0.03084 | 0.01528 |
| O 0 | 0.17410 | −0.01050 | 0.07703 | −0.12824 | 0.00665 |
| O 1 | 0.00308 | 0.00563 | 0.00394 | 0.00203 | 0.05826 |
| O 2 | −0.00238 | −0.00157 | −0.00085 | −0.06579 | 0.00775 |
| H 0 | 0.08534 | 0.22976 | −0.16526 | −0.02341 | −0.20227 |
| H 1 | −0.01709 | 0.08553 | 0.56071 | 0.32296 | −0.04220 |
| H 2 | −0.02266 | −0.04607 | 0.21012 | 0.11684 | −0.09207 |
| H 0 | 0.07137 | 0.27300 | −0.05492 | −0.01664 | −0.67550 |
| H 1 | −0.20974 | −0.50273 | 0.16166 | −0.15387 | −0.09172 |
| H 2 | −0.43345 | 0.20634 | 0.23450 | −0.53572 | −0.24342 |
| H 0 | −0.11270 | −0.40561 | −0.10367 | −0.28659 | −0.25799 |
| H 1 | 0.19697 | −0.32234 | −0.12908 | 0.32587 | −0.50328 |
| H 2 | 0.10362 | −0.08816 | 0.36724 | 0.16776 | −0.14589 |

| freq. | 1472.82 | 1347.75 | 976.30 | 953.34 | 726.92 |
|---|---|---|---|---|---|
| C 0 | −0.00035 | −0.00041 | −0.00023 | 0.00041 | 0.00022 |
| C 1 | −0.00023 | 0.00048 | 0.00058 | −0.00025 | −0.00027 |
| C 2 | −0.00012 | 0.00010 | 0.00013 | 0.00023 | −0.00001 |
| S 0 | 0.00777 | −0.03896 | −0.10126 | 0.06510 | 0.04066 |
| S 1 | 0.00300 | −0.00002 | 0.00015 | −0.00034 | 0.00023 |
| S 2 | −0.00058 | −0.00097 | −0.00207 | 0.00189 | 0.00118 |
| N 0 | −0.00710 | 0.00159 | 0.01057 | −0.00039 | −0.00308 |
| N 1 | 0.00882 | 0.00173 | 0.00647 | 0.00569 | 0.00052 |
| N 2 | 0.02739 | 0.00927 | −0.00130 | 0.00075 | −0.00081 |
| O 0 | −0.05399 | −0.00239 | 0.00354 | 0.00255 | −0.00035 |
| O 1 | 0.00783 | 0.03234 | −0.14634 | −0.13796 | −0.03819 |
| O 2 | 0.16060 | 0.04965 | 0.01144 | −0.10870 | 0.04547 |
| H 0 | 0.07784 | −0.62177 | 0.06113 | −0.23172 | 0.51992 |
| H 1 | −0.06650 | −0.34040 | 0.02429 | −0.29684 | −0.10327 |
| H 2 | −0.05790 | 0.00151 | 0.02539 | −0.10183 | −0.13663 |
| H 0 | 0.07263 | 0.20334 | 0.12792 | −0.08399 | −0.27822 |
| H 1 | −0.12308 | 0.08007 | 0.30525 | −0.15078 | 0.37982 |
| H 2 | −0.19891 | 0.14445 | −0.27066 | 0.06805 | 0.14629 |
| H 0 | 0.18741 | −0.50758 | −0.11375 | 0.00125 | −0.51971 |
| H 1 | −0.21786 | 0.21520 | −0.33266 | −0.14401 | 0.23977 |
| H 2 | 0.60386 | 0.10778 | −0.10035 | 0.39949 | 0.11404 |

| freq. | 595.21 | 307.65 | −194.48 | 200.98 | 146.93 |
|---|---|---|---|---|---|
| C 0 | 0.00033 | −0.00030 | 0.09404 | 0.11708 | −0.07617 |
| C 1 | −0.00053 | 0.00052 | −0.07542 | −0.09374 | 0.05572 |
| C 2 | −0.00018 | −0.00005 | 0.10256 | 0.11886 | −0.08185 |
| S 0 | 0.09067 | −0.07153 | 0.00439 | −0.00119 | 0.00535 |
| S 1 | −0.00000 | 0.00001 | 0.05134 | −0.08881 | −0.06396 |
| S 2 | 0.00247 | −0.00247 | −0.03198 | −0.01056 | −0.07100 |
| N 0 | −0.00769 | 0.00524 | 0.10190 | −0.03561 | 0.08788 |
| N 1 | −0.00087 | −0.00511 | 0.15523 | −0.06016 | 0.05473 |
| N 2 | −0.00146 | 0.00417 | 0.04844 | 0.02950 | 0.08038 |
| O 0 | −0.00128 | −0.00036 | 0.03517 | −0.00296 | 0.00316 |
| O 1 | 0.05283 | 0.11312 | 0.01018 | −0.00116 | 0.00842 |
| O 2 | −0.02024 | −0.12038 | −0.00696 | −0.00434 | −0.01519 |
| H 0 | −0.03789 | 0.29001 | −0.01275 | 0.04810 | −0.12601 |
| H 1 | −0.05398 | −0.27433 | 0.01418 | 0.14131 | 0.37133 |
| H 2 | 0.12872 | −0.06604 | 0.11143 | −0.33443 | −0.28294 |
| H 0 | 0.46394 | 0.04125 | −0.08458 | 0.18441 | 0.00711 |
| H 1 | 0.44414 | 0.15556 | −0.00199 | −0.14297 | 0.12919 |
| H 2 | −0.30192 | 0.05214 | −0.13379 | 0.07554 | 0.08140 |
| H 0 | -0.12168 | 0.01416 | -0.02389 | -0.13015 | -0.03797 |
| H 1 | -0.35704 | -0.15025 | 0.05900 | -0.09424 | -0.03614 |
| H 2 | -0.10894 | 0.42418 | 0.02979 | -0.02574 | 0.10577 |

CH3S˙ radical:
UCCSD(T)-F12A/VQZ-F12 E = -437.574724 Hartree

| | | | |
|---|---|---|---|
| S | 0.0000000000 | -0.0009647639 | -0.5960679814 |
| C | 0.0000000000 | -0.0050463963 | 1.1972540056 |
| H | 0.0000000000 | 1.0440137669 | 1.5060504727 |
| H | 0.8967959193 | -0.4765961457 | 1.5932064950 |
| H | -0.8967959193 | -0.4765961457 | 1.5932064950 |

NO:
UCCSD(T)-F12A/VQZ-F12 E = -129.778072 Hartree

| | | | |
|---|---|---|---|
| N | 0.0000000000 | 0.0000000000 | -0.6138467885 |
| O | 0.0000000000 | 0.0000000000 | 0.5376867885 |

CCSD-F12a/VQZ-F12a Geometries and Energies

Cis-CH3SNO:
CCSD-F12A/VQZ-F12 E = -567.359266 Hartree

| | | | |
|---|---|---|---|
| C | -1.0796612242 | 0.0000000000 | -1.2179254480 |
| S | 0.6251014081 | 0.0000000000 | -0.6724586834 |
| N | 0.4868205128 | 0.0000000000 | 1.0870945947 |
| O | -0.5995785782 | 0.0000000000 | 1.5576590538 |
| H | -1.2814962660 | -0.8911909822 | -1.8031282763 |
| H | -1.7019354918 | 0.0000000000 | -0.3232334742 |
| H | -1.2814962660 | 0.8911909822 | -1.8031282763 |

Trans-CH3SNO:
CCSD-F12A/VQZ-F12 E = -567.358062 Hartree

| | | | |
|---|---|---|---|
| C | 0.0000000000 | 0.6333077488 | -1.7863079879 |
| S | 0.0000000000 | -0.6005385058 | -0.4828810875 |
| N | 0.0000000000 | 0.5562413465 | 0.8608347898 |
| O | 0.0000000000 | 0.0727927560 | 1.9391317949 |
| H | -0.8894750411 | 0.5315323450 | -2.3991849864 |
| H | 0.0000000000 | 1.6065972309 | -1.2991030736 |
| H | 0.8894750411 | 0.5315323450 | -2.3991849864 |

Isomerization TS:
CCSD-F12A/VQZ-F12 E = -567.340521 Hartree

| | | | |
|---|---|---|---|
| C | -1.0156195115 | 0.0306768998 | -1.3517161122 |
| S | 0.6451698802 | -0.0872973519 | -0.6611634634 |
| N | 0.0731323379 | 0.5296695648 | 1.0367982786 |
| O | -0.3415487498 | -0.3215062387 | 1.7178353702 |
| H | -1.6797117764 | -0.6958148554 | -0.8920587784 |
| H | -1.4081707522 | 1.0376753234 | -1.2335833326 |
| H | -0.9255588045 | -0.1878065779 | -2.4124204832 |

Vibrational frequencies, cm-1
3183.03
3173.83
3070.34
1643.72
1493.63
1478.64
1347.15
980.46
957.44
738.78
609.68
322.47

191.51
145.79
-257.40

ZPE, [1/CM]:  9668.23
ZPE, [H]:  0.044052

Normal Modes

| | 1 A | 2 A | 3 A | 4 A | 5 A |
|---|---|---|---|---|---|
| Wavenumbers [cm-1] | 145.79 | 191.51 | 322.47 | 609.68 | 738.78 |
| Intensities [km/mol] | 0.67 | 1.97 | 111.87 | 82.82 | 1.74 |
| Intensities [relative] | 0.18 | 0.52 | 29.47 | 21.82 | 0.46 |
| CX1 | -0.00846 | -0.01120 | -0.00059 | -0.00446 | 0.19682 |
| CY1 | -0.00957 | 0.02012 | -0.00109 | -0.00676 | -0.01255 |
| CZ1 | 0.06804 | 0.16004 | 0.00562 | -0.00565 | 0.09752 |
| SX2 | 0.02503 | 0.05957 | -0.04777 | -0.00254 | -0.08866 |
| SY2 | -0.01835 | -0.00462 | 0.00610 | -0.04180 | 0.00465 |
| SZ2 | -0.01098 | -0.01359 | 0.11130 | -0.00329 | -0.04270 |
| NX3 | -0.00676 | -0.07960 | 0.05441 | -0.03341 | -0.00326 |
| NY3 | 0.00662 | 0.00658 | 0.01363 | 0.15574 | 0.00441 |
| NZ3 | -0.03181 | -0.06972 | -0.12299 | 0.14627 | 0.00423 |
| OX4 | -0.03457 | -0.03456 | 0.04424 | 0.04060 | -0.00003 |
| OY4 | 0.03299 | -0.01450 | -0.02337 | -0.05074 | -0.00128 |
| OZ4 | -0.01833 | -0.07143 | -0.11688 | -0.11505 | -0.00021 |
| HX5 | 0.12053 | -0.00163 | -0.01989 | -0.04238 | 0.13827 |
| HY5 | -0.32818 | 0.15661 | -0.01665 | 0.00273 | -0.01269 |
| HZ5 | -0.25380 | 0.39363 | -0.05053 | -0.04093 | 0.01661 |
| HX6 | -0.16656 | 0.08933 | -0.00933 | 0.02045 | 0.15360 |
| HY6 | -0.12675 | 0.07302 | -0.00474 | 0.00038 | -0.01891 |
| HZ6 | 0.56818 | 0.04663 | -0.00168 | 0.02576 | 0.02789 |
| HX7 | -0.00653 | -0.19431 | 0.09743 | -0.02427 | 0.22860 |
| HY7 | 0.53712 | -0.18360 | 0.02177 | 0.04814 | -0.00780 |
| HZ7 | -0.04277 | 0.18776 | 0.00959 | -0.01919 | 0.09627 |

| | 6 A | 7 A | 8 A | 9 A | 10 A |
|---|---|---|---|---|---|
| Wavenumbers [cm-1] | 957.44 | 980.46 | 1347.15 | 1478.64 | 1493.63 |
| Intensities [km/mol] | 3.46 | 3.94 | 7.46 | 10.46 | 11.25 |
| Intensities [relative] | 0.91 | 1.04 | 1.97 | 2.75 | 2.96 |
| CX1 | -0.06379 | -0.02626 | -0.09257 | -0.01479 | 0.03319 |
| CY1 | 0.04093 | -0.10145 | 0.00498 | -0.05469 | -0.02277 |
| CZ1 | 0.09281 | 0.03064 | -0.05141 | 0.01594 | -0.04902 |
| SX2 | 0.02126 | 0.00770 | -0.00549 | -0.00043 | 0.00109 |
| SY2 | -0.01042 | 0.02255 | -0.00069 | -0.00326 | -0.00088 |
| SZ2 | -0.01714 | -0.00592 | -0.00580 | 0.00036 | -0.00304 |
| NX3 | -0.00931 | -0.00227 | 0.00153 | 0.00192 | -0.00026 |
| NY3 | 0.00029 | -0.00640 | 0.01179 | 0.00405 | -0.00088 |
| NZ3 | -0.00053 | -0.00558 | 0.00110 | -0.00315 | 0.00075 |
| OX4 | 0.00075 | -0.00041 | -0.00334 | -0.00225 | -0.00007 |
| OY4 | 0.00221 | 0.00072 | -0.00761 | -0.00422 | 0.00050 |
| OZ4 | -0.00439 | 0.00471 | 0.00528 | 0.00423 | -0.00041 |
| HX5 | -0.16595 | -0.56656 | 0.36250 | 0.16211 | -0.30426 |
| HY5 | -0.11681 | 0.22764 | -0.12911 | 0.06852 | 0.50779 |
| HZ5 | -0.30715 | -0.22851 | 0.34757 | 0.41307 | 0.33192 |
| HX6 | -0.28269 | 0.57000 | 0.41117 | -0.04051 | -0.43908 |
| HY6 | 0.00381 | 0.12650 | 0.13798 | 0.01624 | -0.23635 |
| HZ6 | -0.34528 | 0.07866 | 0.31516 | -0.50913 | 0.36346 |
| HX7 | 0.64986 | 0.10266 | 0.53588 | 0.07760 | 0.31792 |
| HY7 | -0.08269 | 0.21503 | -0.08905 | 0.68123 | 0.03214 |
| HZ7 | 0.16856 | -0.02413 | 0.03540 | -0.12890 | -0.01847 |

| | 11 A | 12 A | 13 A | 14 A |
|---|---|---|---|---|
| Wavenumbers [cm-1] | 1643.72 | 3070.34 | 3173.83 | 3183.03 |
| Intensities [km/mol] | 379.65 | 15.34 | 2.65 | 1.48 |
| Intensities [relative] | 100.00 | 4.04 | 0.70 | 0.39 |
| CX1 | 0.00333 | 0.04115 | -0.00659 | 0.03463 |
| CY1 | 0.00224 | -0.00640 | -0.08988 | 0.00172 |
| CZ1 | 0.00376 | 0.01327 | -0.00312 | -0.08182 |
| SX2 | 0.00549 | -0.00002 | 0.00003 | -0.00008 |
| SY2 | -0.00412 | 0.00010 | 0.00011 | -0.00002 |
| SZ2 | -0.01084 | 0.00025 | -0.00006 | -0.00007 |
| NX3 | 0.06087 | 0.00041 | 0.00013 | 0.00048 |
| NY3 | 0.15399 | 0.00007 | 0.00020 | 0.00050 |
| NZ3 | -0.09271 | -0.00043 | 0.00015 | -0.00035 |
| OX4 | -0.06181 | -0.00015 | -0.00004 | -0.00024 |
| OY4 | -0.12702 | -0.00021 | -0.00026 | -0.00046 |
| OZ4 | 0.10215 | 0.00023 | 0.00000 | 0.00048 |
| HX5 | -0.02767 | -0.33964 | 0.35131 | -0.29557 |
| HY5 | 0.00758 | -0.39645 | 0.38178 | -0.34442 |
| HZ5 | -0.03470 | 0.25401 | -0.25887 | 0.20342 |
| HX6 | -0.02440 | -0.20601 | -0.25424 | -0.05316 |
| HY6 | -0.01054 | 0.57501 | 0.66448 | 0.16255 |
| HZ6 | 0.00006 | 0.07271 | 0.08410 | 0.00240 |
| HX7 | -0.02711 | 0.05258 | -0.02082 | -0.06412 |
| HY7 | -0.01618 | -0.10300 | 0.02246 | 0.16238 |
| HZ7 | 0.00151 | -0.49047 | 0.21177 | 0.76870 |

Normal Modes of imaginary frequencies

| | 1 |
|---|---|
| Wavenumbers [cm-1] | 257.40 |
| Intensities [km/mol] | 1.17 |
| Intensities [relative] | 0.31 |
| CX1 | -0.02611 |
| CY1 | -0.06824 |
| CZ1 | 0.03637 |
| SX2 | 0.00058 |
| SY2 | 0.03706 |
| SZ2 | -0.01269 |
| NX3 | 0.17202 |
| NY3 | -0.01986 |
| NZ3 | 0.06977 |
| OX4 | -0.12671 |
| OY4 | 0.01162 |
| OZ4 | -0.07025 |
| HX5 | 0.00932 |
| HY5 | -0.06470 |
| HZ5 | 0.09258 |
| HX6 | -0.05811 |
| HY6 | -0.07472 |
| HZ6 | -0.02529 |
| HX7 | -0.03772 |
| HY7 | -0.13445 |
| HZ7 | 0.04846 |

$CH_3S^{\cdot}$ radical:

UCCSD-F12A/VQZ-F12 E = -437.558269 Hartree

| | | | |
|---|---|---|---|
| S | 0.0000000000 | -0.0009252121 | -0.5952180905 |
| C | 0.0000000000 | -0.0047686917 | 1.1953902660 |
| H | 0.0000000000 | 1.0410130951 | 1.5069851998 |
| H | 0.8943704493 | -0.4773794486 | 1.5903272193 |
| H | -0.8943704493 | -0.4773794486 | 1.5903272193 |

NO:

UCCSD-F12A/VQZ-F12 E = -129.756697 Hartree

| | | | |
|---|---|---|---|
| N | 0.0000000000 | 0.0000000000 | -0.6090328488 |
| O | 0.0000000000 | 0.0000000000 | 0.5328728488 |

DCSD-F12a/VQZ-F12a Geometries and Energies

Cis-CH3SNO:
DCSD-F12A/VQZ-F12 E = -567.388025 Hartree

| | | | |
|---|---|---|---|
| C | 1.5529922984 | 0.0392378460 | 0.0000000000 |
| S | -0.0015319185 | 0.9300634725 | 0.0000000000 |
| H | 2.1216330811 | 0.2842270182 | 0.8925275614 |
| H | 1.3082157033 | -1.0252031941 | 0.0000000000 |
| H | 2.1216330811 | 0.2842270182 | -0.8925275614 |
| N | -1.2094557695 | -0.3857140560 | 0.0000000000 |
| O | -0.8044804760 | -1.5049481048 | 0.0000000000 |

Trans-CH3SNO:
DCSD-F12A/VQZ-F12 E = -567.386718 Hartree

| | | | |
|---|---|---|---|
| C | 1.7613419797 | 0.4160220575 | 0.0000000000 |
| S | -0.0031064674 | 0.7584473571 | 0.0000000000 |
| H | 2.2238925126 | 0.8311068251 | 0.8907321282 |
| H | 1.8695482587 | -0.6687378833 | 0.0000000000 |
| H | 2.2238925126 | 0.8311068251 | -0.8907321282 |
| N | -0.5046715958 | -0.9626507975 | 0.0000000000 |
| O | -1.6777032004 | -1.1496563841 | 0.0000000000 |

Isomerization TS:
DCSD-F12A/VQZ-F12 E = -567.368908 Hartree

| | | | |
|---|---|---|---|
| C | -1.0349611219 | 0.0295938215 | -1.3361917642 |
| S | 0.6464062938 | -0.0858869786 | -0.6849532010 |
| N | 0.0746965407 | 0.5338771468 | 1.0523108779 |
| O | -0.3263799347 | -0.3269327250 | 1.7358367048 |
| H | -1.6860687465 | -0.7022841317 | -0.8632656604 |
| H | -1.4282829561 | 1.0359466377 | -1.2033928092 |
| H | -0.9704516134 | -0.1838960992 | -2.4010178259 |

Vibrational frequencies, cm-1
3051.13
3152.58
3140.35
1663.97
1493.32
1478.32
1356.63
981.46
960.94
732.25
609.41
327.29
206.79
150.74
-196.25

ZPE, [1/CM]: 9652.59
ZPE, [H]:  0.043980

Normal modes:

| freq. | 3051.13 | 3152.58 | 3140.35 | 1663.97 | 1493.32 |
|---|---|---|---|---|---|
| C 0 | -0.04166 | 0.00665 | -0.01164 | -0.00008 | -0.00011 |
| C 1 | 0.03196 | -0.00305 | -0.08263 | -0.00013 | -0.00004 |
| C 2 | 0.00840 | 0.08976 | -0.00144 | -0.00004 | -0.00014 |
| S 0 | -0.00206 | -0.00118 | -0.00208 | -0.00423 | 0.00328 |
| S 1 | 0.01946 | -0.01421 | -0.03085 | 0.00109 | -0.00090 |
| S 2 | -0.00900 | -0.03380 | 0.01026 | -0.00047 | -0.00313 |
| N 0 | -0.08807 | 0.00492 | -0.04547 | -0.00800 | -0.00075 |
| N 1 | 0.02390 | 0.09324 | −0.02980 | −0.01144 | −0.03393 |
| N 2 | −0.05708 | 0.03902 | 0.08699 | 0.03247 | −0.01634 |
| O 0 | 0.17178 | −0.01044 | 0.08096 | −0.12684 | 0.00625 |
| O 1 | −0.00456 | −0.00584 | −0.00484 | −0.00076 | −0.05872 |
| O 2 | −0.00128 | −0.00162 | 0.00003 | −0.06790 | 0.00748 |
| H 0 | 0.02476 | −0.08419 | −0.56042 | −0.33136 | 0.04806 |
| H 1 | 0.08986 | 0.22917 | −0.17097 | −0.03020 | −0.20331 |
| H 2 | −0.02351 | −0.04008 | 0.19173 | 0.10942 | −0.09730 |
| H 0 | −0.12279 | −0.40694 | 0.08868 | −0.01724 | 0.62168 |
| H 1 | 0.15248 | 0.45128 | −0.11547 | 0.11654 | 0.26292 |
| H 2 | −0.29164 | 0.36013 | 0.27026 | −0.24530 | −0.04873 |
| H 0 | −0.39739 | 0.04853 | 0.15176 | −0.62826 | 0.03702 |
| H 1 | 0.14845 | 0.00505 | 0.35482 | 0.22864 | 0.03067 |
| H 2 | 0.03640 | 0.38143 | −0.03835 | 0.04266 | 0.63405 |

| freq. | 1478.32 | 1356.63 | 981.46 | 960.94 | 732.25 |
|---|---|---|---|---|---|
| C 0 | −0.00035 | −0.00038 | −0.00019 | 0.00040 | 0.00021 |
| C 1 | −0.00022 | 0.00047 | 0.00054 | −0.00023 | −0.00026 |
| C 2 | 0.00011 | −0.00008 | −0.00015 | −0.00021 | 0.00000 |
| S 0 | 0.00818 | −0.03995 | −0.10110 | 0.06448 | 0.04186 |
| S 1 | −0.00300 | −0.00004 | −0.00018 | 0.00033 | −0.00023 |
| S 2 | 0.00057 | 0.00103 | 0.00215 | −0.00191 | −0.00126 |
| N 0 | −0.00720 | 0.00171 | 0.01067 | −0.00041 | −0.00321 |
| N 1 | 0.00947 | 0.00224 | 0.00644 | 0.00579 | 0.00039 |
| N 2 | −0.02679 | −0.00982 | 0.00112 | −0.00077 | 0.00064 |
| O 0 | −0.05729 | −0.00286 | 0.00457 | 0.00393 | −0.00013 |
| O 1 | −0.00948 | −0.03208 | 0.14644 | 0.13953 | 0.03769 |
| O 2 | 0.15933 | 0.05062 | 0.01360 | −0.10674 | 0.04644 |
| H 0 | 0.06223 | 0.34261 | −0.02908 | 0.30069 | 0.10336 |
| H 1 | 0.08009 | −0.61784 | 0.06792 | −0.23221 | 0.52088 |
| H 2 | −0.05136 | 0.00012 | 0.02667 | −0.09882 | −0.12471 |
| H 0 | −0.10041 | −0.18625 | −0.03841 | 0.03657 | 0.35839 |
| H 1 | 0.06880 | −0.09771 | −0.33896 | 0.17131 | −0.25057 |
| H 2 | −0.26855 | 0.43595 | −0.11337 | 0.04841 | 0.47470 |
| H 0 | 0.18778 | −0.36784 | 0.01264 | 0.14352 | −0.38749 |
| H 1 | 0.62113 | 0.11822 | 0.01705 | 0.40092 | 0.12196 |
| H 2 | −0.01750 | −0.00296 | 0.43032 | −0.02573 | −0.01414 |

| freq. | 609.41 | 327.29 | 206.79 | −196.25 | 150.74 |
|---|---|---|---|---|---|
| C 0 | 0.00029 | −0.00028 | 0.09660 | 0.11577 | −0.07558 |
| C 1 | −0.00050 | 0.00049 | −0.07728 | −0.09274 | 0.05523 |
| C 2 | 0.00018 | 0.00004 | −0.10479 | −0.11709 | 0.08069 |
| S 0 | 0.09026 | −0.07151 | 0.00437 | −0.00113 | 0.00555 |
| S 1 | 0.00003 | −0.00003 | −0.05224 | 0.08978 | 0.06159 |
| S 2 | −0.00252 | 0.00251 | 0.03043 | 0.01128 | 0.07172 |
| N 0 | −0.00775 | 0.00530 | 0.09937 | −0.03520 | 0.09031 |
| N 1 | −0.00084 | −0.00519 | 0.15434 | −0.06020 | 0.05966 |
| N 2 | 0.00169 | −0.00407 | −0.04493 | −0.03099 | −0.08020 |
| O 0 | −0.00153 | −0.00087 | 0.03522 | −0.00337 | 0.00383 |
| O 1 | −0.05201 | −0.11124 | −0.01037 | 0.00125 | −0.00852 |
| O 2 | −0.02170 | −0.12127 | −0.00689 | −0.00420 | −0.01529 |
| H 0 | 0.05158 | 0.27633 | −0.01295 | −0.13466 | −0.36345 |
| H 1 | −0.04174 | 0.29408 | −0.01174 | 0.05286 | −0.12195 |
| H 2 | 0.13032 | −0.05726 | 0.11472 | −0.34006 | −0.28882 |
| H 0 | −0.33091 | 0.00110 | 0.08130 | −0.22169 | 0.02597 |
| H 1 | −0.56659 | −0.16007 | 0.01587 | 0.10227 | −0.11645 |
| H 2 | −0.11662 | 0.06257 | −0.09058 | 0.12599 | 0.10342 |
| H 0 | 0.01081 | 0.15339 | −0.11461 | 0.01492 | 0.04597 |
| H 1 | 0.01539 | 0.42234 | 0.04267 | 0.00000 | 0.10005 |
| H 2 | 0.46425 | −0.02848 | 0.00860 | 0.11132 | −0.01251 |

CH3S˙ radical:
UDCSD−F12A/VQZ−F12 E = −437.567824 Hartree

| | | | |
|---|---|---|---|
| S | 0.0000000000 | −0.0009426564 | −0.5960342220 |

```
C          0.0000000000       -0.0049001929        1.1971959626
H          0.0000000000        1.0427912404        1.5073254822
H          0.8958696021       -0.4772075830        1.5923779211
H         -0.8958696021       -0.4772075830        1.5923779211
```

NO:
UDCSD-F12/VQZ-F12 E = -129.767401 Hartree

```
N          0.0000000000        0.0000000000       -0.6130140446
O          0.0000000000        0.0000000000        0.5368540446
```

DCSD/VQZ-F12a Geometries and Energies

Cis-CH3SNO:
DCSD/VQZ-F12 E = -567.358226 Hartree

```
C         -1.0950982866        0.0000000000       -1.2028763746
S          0.6233839266        0.0000000000       -0.6908518041
N          0.4995920813        0.0000000000        1.0954414545
O         -0.5923088454        0.0000000000        1.5714723818
H         -1.3108093030       -0.8929043318       -1.7839401764
H         -1.6976002268        0.0000000000       -0.2911569625
H         -1.3108093030        0.8929043318       -1.7839401764
```

Trans-CH3SNO:
DCSD/VQZ-F12 E = -567.356938 Hartree

```
C          0.0000000000        0.6491308666       -1.7831577953
S          0.0000000000       -0.6079808476       -0.4963431163
N          0.0000000000        0.5580837034        0.8700220363
O          0.0000000000        0.0704100120        1.9542329538
H         -0.8911105391        0.5572753809       -2.3984595267
H          0.0000000000        1.6154988257       -1.2772756310
H          0.8911105391        0.5572753809       -2.3984595267
```

Isomerization TS:
DCSD/VQZ-F12 E = -567.339289 Hartree

```
C         -1.0378057099        0.0296706630       -1.3349447787
S          0.6470727503       -0.0859288249       -0.6874474636
N          0.0755702044        0.5344471775        1.0540070063
O         -0.3256162048       -0.3274814936        1.7380258473
H         -1.6873580935       -0.7050306722       -0.8629361654
H         -1.4324599362        1.0354936139       -1.1971718183
H         -0.9765500750       -0.1794917314       -2.4014108156
```

Vibrational frequencies, cm-1
3047.97
3148.19
3139.24
1659.15
1492.87
1478.55
1355.72
979.81
959.18
730.49
607.68
325.95
206.09
149.34
-194.15

ZPE, [1/CM]:  9640.12
ZPE, [H]:  0.043924

Normal modes:

| freq. | 3047.97 | 3148.19 | 3139.24 | 1659.15 | 1492.87 |
|---|---|---|---|---|---|
| C 0 | -0.04173 | 0.00680 | -0.01146 | -0.00007 | -0.00011 |
| C 1 | 0.03207 | 0.00190 | -0.08267 | -0.00012 | -0.00005 |
| C 2 | -0.00674 | -0.08979 | -0.00313 | 0.00003 | 0.00014 |
| S 0 | -0.00212 | -0.00120 | -0.00211 | -0.00421 | 0.00326 |
| S 1 | -0.01932 | 0.01415 | 0.03101 | -0.00107 | 0.00089 |
| S 2 | 0.00894 | 0.03384 | -0.01026 | 0.00045 | 0.00312 |
| N 0 | -0.08823 | 0.00495 | -0.04514 | -0.00804 | -0.00075 |
| N 1 | 0.02340 | 0.09361 | -0.02894 | -0.01121 | -0.03409 |
| N 2 | 0.05684 | -0.03803 | -0.08752 | -0.03234 | 0.01597 |
| O 0 | 0.17206 | -0.01056 | 0.08038 | -0.12696 | 0.00627 |
| O 1 | -0.00466 | -0.00579 | -0.00495 | -0.00071 | -0.05863 |
| O 2 | -0.00133 | -0.00159 | 0.00031 | -0.06759 | 0.00736 |
| H 0 | 0.08996 | 0.23077 | -0.17239 | -0.03015 | -0.20452 |
| H 1 | -0.02417 | 0.08439 | 0.56272 | 0.33204 | -0.04900 |
| H 2 | -0.02321 | -0.04208 | 0.18563 | 0.10533 | -0.09799 |
| H 0 | 0.07932 | 0.04280 | -0.03029 | 0.13567 | 0.63029 |
| H 1 | 0.12413 | 0.74990 | -0.07459 | 0.06155 | -0.12751 |
| H 2 | 0.35727 | -0.12560 | -0.40191 | 0.32388 | 0.05133 |
| H 0 | -0.15410 | 0.04854 | -0.25818 | -0.14019 | 0.12062 |
| H 1 | 0.36106 | -0.08570 | 0.00418 | 0.61337 | -0.08525 |
| H 2 | -0.01858 | 0.23709 | 0.10005 | 0.02675 | 0.65103 |

| freq. | 1478.55 | 1355.72 | 979.81 | 959.18 | 730.49 |
|---|---|---|---|---|---|
| C 0 | -0.00033 | -0.00038 | -0.00020 | 0.00041 | 0.00021 |
| C 1 | -0.00021 | 0.00047 | 0.00053 | -0.00024 | -0.00027 |
| C 2 | -0.00012 | 0.00010 | 0.00018 | 0.00020 | -0.00002 |
| S 0 | 0.00816 | -0.03993 | -0.10110 | 0.06450 | 0.04182 |
| S 1 | 0.00299 | 0.00004 | 0.00018 | -0.00032 | 0.00022 |
| S 2 | -0.00056 | -0.00105 | -0.00221 | 0.00195 | 0.00128 |
| N 0 | -0.00715 | 0.00178 | 0.01079 | -0.00049 | -0.00325 |
| N 1 | 0.00919 | 0.00211 | 0.00641 | 0.00574 | 0.00043 |
| N 2 | 0.02694 | 0.00985 | -0.00105 | 0.00086 | -0.00068 |
| O 0 | -0.05707 | -0.00285 | 0.00463 | 0.00402 | -0.00013 |
| O 1 | -0.00951 | -0.03193 | 0.14636 | 0.13968 | 0.03761 |
| O 2 | 0.15941 | 0.05015 | 0.01366 | -0.10695 | 0.04636 |
| H 0 | 0.08002 | -0.61705 | 0.06765 | -0.23079 | 0.51961 |
| H 1 | -0.06526 | -0.34468 | 0.02948 | -0.30068 | -0.10387 |
| H 2 | -0.05088 | 0.00548 | 0.02679 | -0.09506 | -0.12441 |
| H 0 | 0.03474 | -0.20402 | -0.25459 | 0.16569 | 0.04509 |
| H 1 | 0.00421 | 0.22096 | -0.12645 | 0.07698 | -0.23575 |
| H 2 | 0.06527 | -0.30935 | 0.15281 | -0.17097 | -0.47246 |
| H 0 | -0.70477 | 0.08123 | 0.01924 | -0.38720 | 0.19104 |
| H 1 | -0.04447 | 0.43565 | -0.02193 | -0.03173 | 0.49281 |
| H 2 | 0.06195 | 0.03874 | 0.45843 | 0.03783 | 0.15233 |

| freq. | 607.68 | 325.95 | -194.15 | 206.09 | 149.34 |
|---|---|---|---|---|---|
| C 0 | 0.00030 | -0.00029 | 0.09612 | 0.11602 | -0.07531 |
| C 1 | -0.00049 | 0.00049 | -0.08222 | -0.09886 | 0.05909 |
| C 2 | -0.00021 | -0.00002 | 0.10057 | 0.11279 | -0.07768 |
| S 0 | 0.09028 | -0.07149 | 0.00445 | -0.00116 | 0.00562 |
| S 1 | -0.00003 | 0.00002 | 0.05207 | -0.08950 | -0.06225 |
| S 2 | 0.00257 | -0.00255 | -0.03056 | -0.01090 | -0.07148 |
| N 0 | -0.00785 | 0.00538 | 0.09974 | -0.03555 | 0.08985 |
| N 1 | -0.00080 | -0.00517 | 0.15430 | -0.06040 | 0.05864 |
| N 2 | -0.00171 | 0.00406 | 0.04612 | 0.03029 | 0.08058 |
| O 0 | -0.00156 | -0.00093 | 0.03522 | -0.00327 | 0.00376 |
| O 1 | -0.05208 | -0.11125 | -0.01040 | 0.00126 | -0.00851 |
| O 2 | -0.02168 | -0.12147 | -0.00690 | -0.00408 | -0.01506 |
| H 0 | -0.04167 | 0.29416 | -0.01219 | 0.05299 | -0.12366 |
| H 1 | -0.05054 | -0.27540 | 0.01469 | 0.13037 | 0.36038 |
| H 2 | 0.13205 | -0.05437 | 0.11392 | -0.34029 | -0.29434 |
| H 0 | −0.61560 | −0.10145 | 0.08090 | −0.08011 | −0.07307 |
| H 1 | −0.06538 | −0.10659 | −0.06318 | 0.27586 | −0.06679 |
| H 2 | 0.20077 | −0.21872 | 0.08145 | −0.03654 | −0.14702 |
| H 0 | −0.00981 | −0.39098 | −0.02564 | 0.00621 | −0.06088 |
| H 1 | −0.01130 | −0.00267 | 0.10979 | −0.02398 | 0.00967 |
| H 2 | 0.48544 | 0.09982 | 0.01710 | 0.05919 | 0.04513 |

CH3S˙ radical:
UDCSD/VQZ−F12 E = −437.553890 Hartree

| | | | |
|---|---|---|---|
| S | 0.0000000000 | −0.0009399386 | −0.5967207732 |
| C | 0.0000000000 | −0.0048842898 | 1.1985022986 |
| H | 0.0000000000 | 1.0432376284 | 1.5095297117 |
| H | 0.8962191682 | −0.4775687538 | 1.5944111272 |
| H | −0.8962191682 | −0.4775687538 | 1.5944111272 |

NO:
UDCSD/VQZ−F12 E = −129.756482 Hartree

| | | | |
|---|---|---|---|
| N | 0.0000000000 | 0.0000000000 | −0.6137825760 |
| O | 0.0000000000 | 0.0000000000 | 0.5376225760 |